# KAMLAND-EXPERIMENT AND SOLITON-LIKE NUCLEAR GEOREACTOR
# PART 1. COMPARISON OF THEORY WITH EXPERIMENT

*V.D. Rusov[1*], D.A. Litvinov[1], S. Cht. Mavrodiev[2], E.P. Linnik[1],*
*V.N. Vaschenko[3], T.N. Zelentsova[1], R. Beglaryan[1], V.A. Tarasov[1],*
*S. Chernegenko[1], V.P. Smolyar[1], P.O. Molchinikolov[1], K.K. Merkotan[1]*

[1]*Odessa National Polytechnic University, Ukraine,*
[2]*Institute for Nuclear Research and Nuclear Energy, BAS, Sofia, Bulgaria,*
[3]*National Antarctic Centre, Kiev, Ukraine*

We give an alternative description of the data produced in the KamLAND experiment, assuming the existence of a natural nuclear reactor on the boundary of the liquid and solid phases of the Earth's core. Analyzing the uncertainty of antineutrino spectrum of georeactor origin, we show that the theoretical (which takes into account the soliton-like nuclear georeactor) total reactor antineutrino spectra describe with good accuracy the experimental KamLAND-data over the years of 2002-2007 and 2002-2009, respectively. At the same time the parameters of mixing ($\Delta m^2_{21}$=2.5· $10^{-5}$ eV$^2$, $\tan^2\theta_{12}$=0.437) calculated within the framework of georeactor hypothesis substantially differ from the parameters of mixing ($\Delta m^2_{21}$=7.49· $10^{-5}$ eV$^2$, $\tan^2\theta_{12}$=0.436) obtained in KamLAND-experiment for total exposure over the period of 2002-2009.

By traingulation of KamLAND and Borexino data we have constructed the coordinate location of soliton-like nuclear georeactors on the boundary of the liquid and solid phases of the Earth core.

Based on the necessary condition of full synchronization of geological (magnetic) time scale and time evolution of heat power of nuclear georeactor, which plays the role of energy source of the Earth magnetic field, and also the strong negative correlation between magnetic field of the solar tachocline zone and magnetic field of the Earth liquid core (Y-component) we have obtain the estimation of nuclear georeactor average heat power ~30 TW over the years 2002-2009.

## 1. Introduction

It is obvious now that the experiments by the KamLAND-collobaration over the last 8 years [1-5] are extremely important not only for observation of reactor antineutrino oscillations, but because they make it possible for the first time to verify one of most vivid and mysterious ideas in nuclear geophysics – the hypothesis of natural nuclear georeactor existence [6-20]. In spite of its singularity and long history, this hypothesis becomes especially attractive today because it enables clearly to explain from the physical standpoint different unrelated, at the first glance, geophysical anomalous phenomena whose fundamental nature is beyond any doubt.

First of all it concerns the problem of $^3$He and $^4$He isotopes origin in the Earth interior, whose concentration ratio, as is well known, "mystically" increases to the center of Earth [21, 22]. This is practically impossible to explain by existing now models of the origin of the anomalous $^3$He concentration and $^3$He/$^4$He ratio distribution in the Earth's interior so long as they have serious contradictions. For example, Anderson et al. has pointed out [23]: "The model whereby high $^3$He/$^4$He is attributed to a lower mantle source, and is thus effectively an indicator of plumes, is becoming increasingly untenable as evidence for a shallow origin for many high $^3$He/$^4$He hot spots accumulates. Shallow, low $^4$He for high $^3$He/$^4$He are logically reasonable, cannot be ruled out, and need to be rigorously tested if we are to be understand the full implications of this important geochemical tracer". Apparently, the most advanced model, which is devoid of the mentioned contradictions, is the so-called Gonnermann-Mukhopadhyay model, preserving noble gases in convective mantle [24]. However this model ignores the possible high concentrations of $^{238}$U and $^{232}$Th in the outer core (as it is shown by numerous laboratory experiments [16, 18, 20]), and this is the weak point of this model. At the same time, it is shown [17] that, if the nuclear georeactor exists, within the framework of model, which takes into account the georeactor thermal power and distribution of $^{238}$U and $^{232}$Th in the Earth's interior, it is possible also to obtain a good description of the known experimental $^3$He/$^4$He distributions in the crust and mantle.

---
[*] Corresponding author: Rusov Vitaliy, E-mail: siiis@te.net.ua

A potent argument in favour of the nuclear georeactor existence are results of recent seismo-tomography researches of the anomalous high heat flow (13± 4 TW) on the core-mantle boundary. This heat is much higher than the radiogenic heat in the lower mantle (*D*"-region) [25]. To explain such an anomalous high heat flow the authors of this paper have advanced the hypothesis of young solid core of the Earth whose the crystallization energy is the cause of anomalous temperature effect.

In full measure it concerns the known problem of nature of an energy source maintaining the convection in the Earth liquid core or, more precisely, the mechanism of magneto-hydrodynamic dynamo generating the Earth magnetic field. It is obvious, that the well-known $^{40}$K-mechanism of radiogenic heat production in the solid core of the Earth does not solve the problem on the whole, because it can not explain the heat flows balance on the core-mantle boundary (see [26] and refs therein) . It should be also mentioned the so-called mechanism of the Earth magnetic field inversions closely associated with the problem of convection in the Earth liquid core. It seems to be strange, but both these fundamental problems have a simple and physically clear solution within the framework of hypothesis of existence of the natural nuclear georeactor on the boundary of the liquid and solid phases of the Earth [17, 27].

If the georeactor hypothesis is true, the fluctuations of georeactor thermal power can influence on Earth's global climate in the form of anomalous temperature jumps in the following way. Strong fluctuations of georeactor thermal power can lead to the partial blocking of convection in the liquid core [27] and the change of an angular velocity of liquid geosphere rotation, thereby, by virtue of a conservation law of Earth's angular moment to the change of angular velocity of mantle and the Earth's surface, respectively. This means that the heat or, more precisely, dissipation energy caused by friction of earthly surface and bottom layer can make a considerable contribution to total energybalance of the atmosphere and thereby significantly to influence on the Earth global climate evolution [27].

However, in spite of obvious attractiveness of this hypothesis there are some difficulties for its perception predetermined by non-trivial properties which georeactor must possess. At first, natural, i.e. unenriched, uranium or thorium must be used as a nuclear fuel. Secondly, traditional control rods are completely absent in the reactivity regulation system of reactor. Thirdly, in spite of the absence of control rods a reactor must possess the property of so-called inner safety. It means that the critical state of reactor core must be permanently maintained in any situation, i.e. the normal operation of reactor is automatically maintained not as a result of operator activity, but by virtue of physical reasons-laws preventing the explosive development of chain reaction by the natural way. Figuratively speaking, the reactor with inner safety is the "nuclear installation which never explode" [28].

It seems to be strange, but reactors satisfying such unusual requirements are possible in reality. For the first time the idea of such a self-regulating fast reactor (so-called mode of breed-and-burn) was expressed in a general form at II Genevan conference in 1958 by Russian physicists Feynberg and Kunegin [29] and relatively recently "reanimated" as an idea of the self-regulating fast reactor in traveling-wave mode of nuclear burning by L. Feoktistov [30] and independently by Teller, Ishikawa and Wood [31].

To interpret the experimental KamLAND antineutrino spectra [3-5] we consider below the properties of such an unusual reactor.

## 2. Soliton-like nuclear georeactor and the KamLAND antineutrino spectra (experiments over the period of 2002-2004)

The main idea of reactor with inner safety consists in selection of fuel composition so that, at first, the characteristic time $\tau_\beta$ of nuclear burning of fuel active (fissile) component is substantially greater than the characteristic time of delayed neutrons production and, secondly, necessary self-regulation conditions are fulfilled during the reactor operation (that always take place when the equilibrium concentration of fuel active component is greater than its critical concentration [30]). These very important conditions can practically always to be attained, if among other reactions in a reactor core the chain of nuclear transformations of the Feoktistov uranium-plutonium cycle type [30]

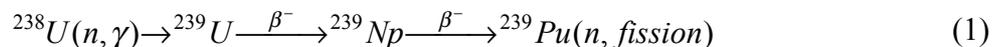

$$^{238}U(n,\gamma) \rightarrow {}^{239}U \xrightarrow{\beta^-} {}^{239}Np \xrightarrow{\beta^-} {}^{239}Pu(n, fission) \qquad (1)$$

or the Teller-Ishikawa-Wood thorium-uranium cycle type [31]

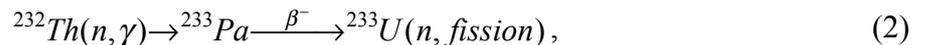

$$^{232}Th(n,\gamma) \rightarrow {}^{233}Pa \xrightarrow{\beta^-} {}^{233}U(n, fission), \qquad (2)$$

will be enough appreciable.

In both cases the active components of nuclear fuel are the generated fissile isotopes of $^{239}$Pu (1) or $^{233}$U (2). The characteristic time of such a reaction, i.e. the respective $\beta$-decay time, is approximately equal to $\tau_\beta =2.3/\ln2 \approx 3.3$ days and $\tau_\beta \approx 39.5$ days for reactions (1) and (2), respectively. This is several orders of magnitude greater than the time of delayed neutrons production.

Self-regulation of nuclear burning process (under indicated above ration between the equilibrium and critical concentrations of fuel active components [30]) takes place because such a system, which is left by itself, can not pass from a critical state to reactor runaway mode as a critical concentration is bounded from above by the finite value of plutonium equilibrium concentration, i.e. $\tilde{n}_{Pu} > n_{crit}$. On phenomenological level the self-regulation of nuclear burning is manifested as follows. Increase of a neutron flux due to some reasons will result in rapid burnup, for example, of plutonium, i.e. in decrease of its concentration, and therefore in decrease of neutron flux, while the new nuclei of $^{239}$Pu are produced with the same rate during $\tau_\beta$ =3.3 days. And vice versa, if the neutron flux is sharply decreased due to external action, the burnup rate decreases too and the plutonium accumulation rate will be increased as well as the number of neutrons produced in a reactor after approximately same time. Analogical situation will be observed for the thorium-uranium cycle (2), but in time $\tau_\beta$ =39.5 days.

Generation of the system of kinetic equations for components of nuclear fuel and neutrons (as a diffusion approximation) in such chains is sufficiently simple and was in detail described in our paper [17]. Typical for such a problem solutions in the form of soliton-like concentration wave of nuclear fuel components and neutrons (Eqs. (3)-(9) in [17]) are shown in Figure 1. Within the framework of soliton-like fast reactor theory it is easy to show that the phase velocity $u$ of nuclear burning is predetermined by following approximate equality [32]

$$\frac{u\tau_\beta}{2L} \simeq \left(\frac{8}{3\sqrt{\pi}}\right)^6 a^4 \exp\left(-\frac{64}{9\pi}a^2\right), \quad a^2 = \frac{\pi^2}{4} \cdot \frac{n_{crit}}{\tilde{n}_{fis} - n_{crit}}, \qquad (3)$$

where $\tilde{n}_{fis}$ and $n_{crit}$ are the equilibrium and critical concentrations of active (fissile) isotope, respectively; $L$ is the average diffusion distance for neutron, $\tau_\beta$ is the delay time caused by active

(fissile) isotope production, which is equal to the effective period of intermediate nuclei $\beta$-decay in the uranium-plutonium cycle (1) or thorium-uranium cycle (2).

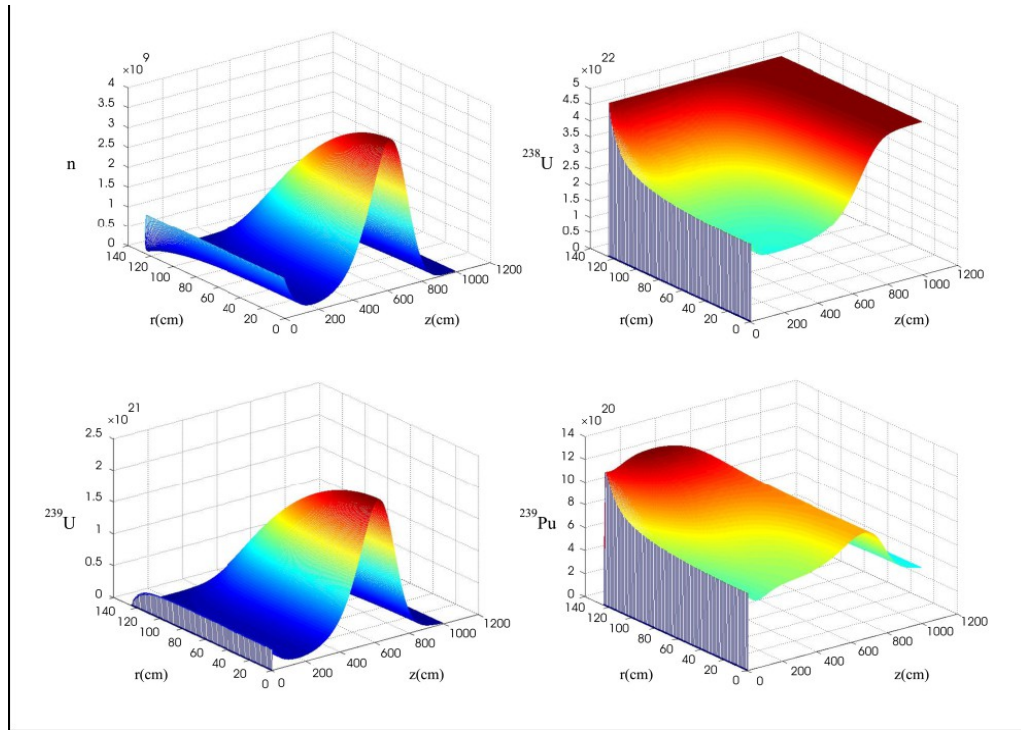

**Fig. 1.** Concentration kinetics of neutrons, $^{238}$U, $^{239}$U, $^{239}$Pu in the core of cylindrical reactor with radius of 125 cm and 1000 cm long at the time of 240 days. Here $r$ is transverse spatial coordinate axis (cylinder radius), $z$ is longitudinal spatial coordinate axis (cylinder length).

Note that Eq. (3) automatically contains the self-regulation condition for nuclear burning because the existence of wave is predetermined by the inequality $\tilde{n}_{fis} > n_{crit}$. In other words, Eq. (3) is necessary physical requirement for the existence of soliton-like neutron wave of nuclear burning. We indicate for a comparison that, as it follows from Eq. (3), the upper bounds of phase velocity of nuclear burning wave are 3.70 cm/day for the uranium-plutonium cycle (1) and 0.31 cm/day for the thorium-uranium cycle (2) at almost equal average diffusion distance ($L\sim5$ cm) for fast neutrons (1 MeV) both for uranium and thorium.

Finally, we consider the some important details and properties of such a soliton-like fast reactor, assuming the existence of which, we have obtained the theoretical spectra of reactor antineutrino and terrestrial antineutrino which are in good agreement with the experimental KamLAND data [17] corresponding to the first [1] and third [3] exposures.

According to our notions, a soliton-lke fast reactor is located on the boundary of the liquid and solid phases of the Earth [17]. The average thickness of such a shell-boundary, which has increased density and mosaic structure, is ~2.2 km [32]. In our opinion, the most advanced mechanism for formation of such a shell below the mantle now are the experimental results of Anisichkin et al. [16, 18] and Hueshao-Secco [34]. According to these results, the chemically stable high-density actinide compounds (particularly uranium carbides and uranium dioxides) lose most of their lithophilic properties at high pressure, sink together with melted iron and concentrate in the Earth's core consequent to the initial gravitational differentiation of the planet. On the other words, during early stages of the evolution of the Earth and other planets, U and Th oxides and carbides (as the most dense, refractory, and marginally soluble at high pressures) accumulated from a magma "ocean" on the solid inner core of the planet, thereby activating chain nuclear reactions, and, in particular, a progressing wave of Feoktisov and/or Teller-Ishikawa-Wood type.

What is the thermal power of such a reactor? As a natural quantitative criterion of the georeactor thermal power we used the well-known (based on the geochemical measurements) $^3He/^4He$ radial distribution in the Earth's interior [17]. It turned out that the experimental average values of $^3He/^4He$ for crust, the depleted upper mantle, the mantle (minus the depleted upper mantle) and the so called *D"*-region in the lower mantle are in good agreement with the theoretical data obtained by the model of Feoktistov's uranium-plutonium georeactor with thermal power of 30 TW [17]. Figure 2 shows the especial experimental investigation of geologically produced antineutrinos with KamLAND [3] and an alternative description of these data by our georeactor model [17].

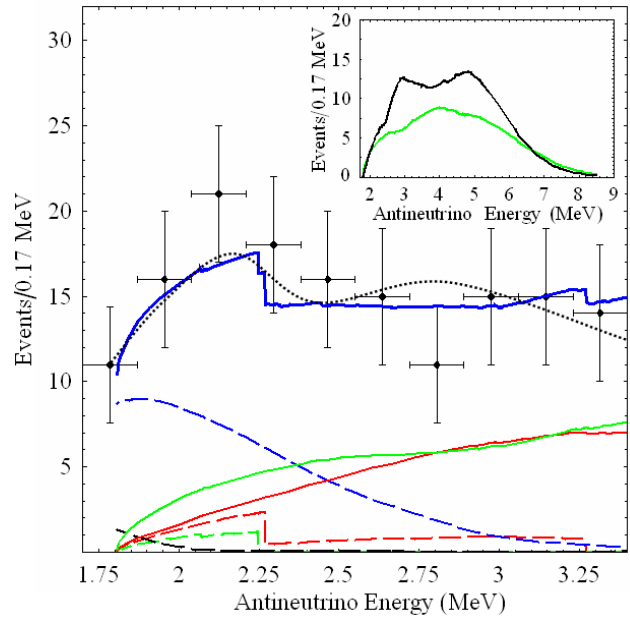

**Fig. 2.** The $\tilde{\nu}_e$ energy spectra in KamLAND [17]. Main panel, experimental points (solid black dots with error bars) together the total expectation obtained in KamLAND experiment (dotted black line) [3] and presented paper (thick solid blue line). Also shown are expected neutrino spectrum (solid green line) from Japan's reactor, the expected neutrino spectrum from georeactor 30 TW (red line), the expected signals from $^{238}U$ (dashed red line) and $^{232}Th$ (dashed green line) geoneutrinos, $^{13}C(\alpha,n)^{16}O$ reactions (dashed blue line) and accidentals (dashed black line). Inset, expected spectra obtained in KamLAND experiment (solid black line) [3] and our paper [17] (solid green line) extended to higher energy.

We need to note that, in spite of the fact that the experimental KamLAND-data are well described within the framework of georeactor model [17] (see Figure 2), some geophysicists have one's doubts about the existence of georeactor, and the value of georeactor power (30 TW) arouses a special mistrust. In this connection we would like to pay attention for the strange restriction ($W \leq$ 6.2 TW) on the value of nuclear georeactor thermal power $W$, which, unfortunately, is frequently met in the scientific literature recently [3, 4, 35, 36], and strongly masks and distorts the clear understanding of problem of georeactor existence, which is intricate enough by itself. Below we ground a complete insolvency of this restriction. One of the conclusions of the KamLAND–colloboration is the upper bound of nuclear georeactor thermal power ($W \leq$ 6.2 TW at 90% C.L.), which is a direct consequence of uncertainty of KamLAND experimental data [4]. However, it is necessary to keep firmly in mind that the restriction of 6.2 TW on georeactor power is true only for the concrete parameters of mixing, i.e. for $\Delta m^2_{21}$

=7.58· 10⁻⁵ eV², tan²$\theta_{12}$=0.56, which takes into account the existence of georeactor within the framework of nonzero hypothesis [4], but absolutely ignores such a nontrivial property of the nuclear georeactor as an uncertainty of georeactor antineutrino spectrum, which in the case of soliton-like nuclear georeactor reaches ~100%. As it will be shown below, the account of this uncertainty within the framework of maximum likelihood function leads (in the minimization of the $\chi^2$-function) to considerable expansion of restriction on the nuclear georeactor heat power (~30 TW) and, accordingly, to the new oscillation parameters ($\Delta m^2_{21}$=2.5· 10⁻⁵ eV², tan²$\theta_{12}$= 0.437) for reactor antineutrino.

Another widespread error is related to determination of the Earth geothermal power $W_{Earth}$. It is known that there are two estimations of $W_{Earth}$, i.e., ~ 33 ± 1 TW [37] and ~ 44 ± 1 TW [38]. We are not going to participate in the discussion among the authors of these estimations concerning necessity of taking into account the hydrothermal circulation. We would like only to emphasize that these estimations are 1.7–2.3 times greater than the radiogenic heat contribution (from the decay $^{238}$U, $^{232}$Th and $^{40}$K in the mantle and crust), which is 19.5 TW [17]. D.L. Anderson [39] refers to this difference as "the missing heat source problem" and summarizes the situation in the following words: "Global heat flow estimates range from 30 to 44 TW… Estimates of the radiogenic contribution (from the decay of U, Th and K in the mantle), based on cosmochemical considerations, vary from 19 to 31 TW. Thus, there is either a good balance between current input and output… or there is a serious missing heat source problem, up to a deficit of 25 TW…" Because of this missing heat, some researchers think that, if a reactor exists, its thermal power must defray the existent deficit of geothermal energy. It is correct, if to keep in mind the thermal power of reactor which operated in the remote past, but which does not operate today. The difference between the heat generated now by a reactor in the Earth interior and the experimentally observed geothermal heat [37, 38] is very significant due to the high thermal inertia of the Earth. In other words, it is necessary to take into account that the heat generated in the Earth interior is not instantly transferred to the surface, but delays (due to a low heat conductivity) in a time of thermal relaxation of the Earth ($\tau_E \approx 10^9$ years) [40, 41]. From here it follows that it is impossible to summarize heat flows which have the different spatial-temporal origin.

## 3. The nonstationary soliton-like nuclear georeactor and KamLAND antineutrino spectrum (experiments over the period of 2002-2007)

Now we consider the use of idea of soliton-like nuclear georeactor to describe the KamLAND experimental antineutrino spectra over the period of 2002-2007 [4]. For this purpose let us estimate an uncertainty of nuclear georeactor thermal power and an uncertainty of georeactor antineutrino spectrum, respectively. Note that, generally speaking, such an uranium-plutonium georeactor can consist of a few tens or hundreds of reactors (with the total thermal power of 30 TW), which represents the individual burning «rivers» and «lakes» of an inhomogeneous actinide shell located in the valleys of rough surface of the Earth's solid core [17]. In the general case, the fission rate of $^{239}Pu$ nuclei for the uranium-plutonium cycle (1) in the one-group approximation can be written down in the form

$$\lambda_{Pu} = \varphi \sigma_f n_{Pu} V, \qquad (4)$$

where $\Phi = \upsilon n$ is the neutron-flux density; $\upsilon$ is the neutron velocity; $n$ is the neutron concentration; $\sigma_f$ is the fission cross-section for $^{239}Pu$; $n_{Pu}$ is the $^{239}Pu$ concentration; $V$ is the volume of burning area.

It is easy to see that due to the random character of critical and equilibrium concentrations of plutonium in an actinoide shell and also a stochastic geometry of the "rivers" and "lakes" of actinoide medium the relative variations of neutron flux density $\phi$, the plutonium concentration $n$ and the volume of burning areas can run up to 50% and more. Then ignoring the possible variations of fission cross-section for plutonium, we can write down the following relation for the relative variation of fission rate:

$$\frac{\Delta \lambda_{Pu}}{\lambda_{Pu}} = \left[\left(\frac{\Delta \varphi}{\varphi}\right)^2 + \left(\frac{\Delta n_{Pu}}{n_{Pu}}\right)^2 + \left(\frac{\Delta V}{V}\right)^2\right]^{1/2} \geq 0.87, \quad \frac{\Delta \sigma}{\sigma} \ll 1. \tag{5}$$

On the other hand, it is obvious that a kinetics of georeactor, which operates on the boundary of the liquid and solid phases of the Earth core at the temperature of 5000-6000 K and pressure of a few hundreds of thousands atmospheres, must necessarily take into account a heat transfer kinetics. This is caused by the fact that under such thermodynamics conditions between these kinetics non-trivial feed-backs can arise, which will significantly change the "traditional" kinetics of neutrons and nuclear reactions. It should be noted that such a problem, apparently, is first solved within the frameworks of reactor physics. We have obtained the dependence of fission cross-section $\langle \sigma_f \rangle$ for the $^{239}_{94}Pu$ nuclei averaged over the neutron spectrum on the nuclear fuel temperature $T$ by the computational experiment with an allowance for the moderation of neutrons and neutron resonance absorption (Figure 3).

This dependence is a power function[1] in the 4000 to 6000 K range (see Figure 3):

$$\langle \sigma_f^{Pu} \rangle \sim T^\alpha, \quad where \quad \alpha \geq 2. \tag{6}$$

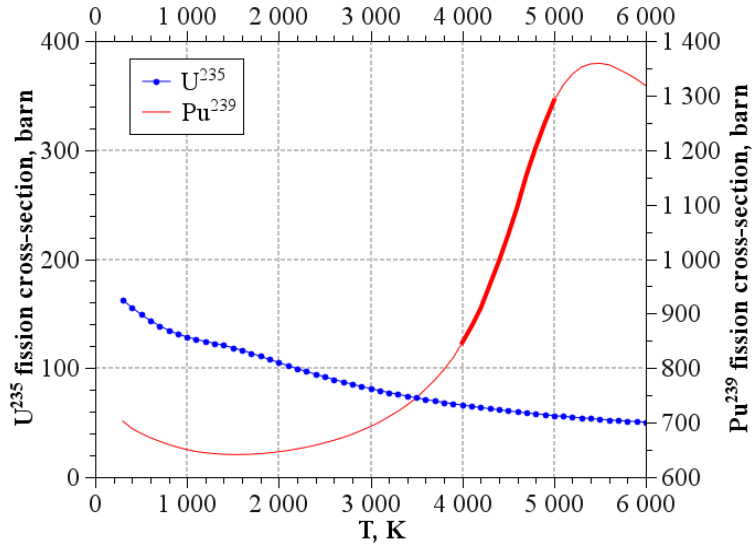

**Figure 3.** Dependence of $^{239}$Pu fission cross-section averaged over the neutron spectrum on fuel medium temperature for limiting energy ($3kT$) of the Fermi and Maxwell spectra. The similar dependence for the $^{235}U$ fission cross-section is shown for comparison.

---

[1] It is interesting, that such a behavior of cross-section on the medium temperature is appropriate for the fission cross-section and capture cross-section section of $^{239}$Pu and absolutely is not appropriate for similar cross-sections of the $^{235}$U nuclei, which have the classical dependence of $1/v$ type.

From Figure 3 follows that the weak variations of temperature in the 4000 to 6000 K range can cause the strong variations of fission cross-section $\langle\sigma_f\rangle$ for $^{239}$Pu, which can run up to 100% and more. In its turn, the variations of fission cross-section $\langle\sigma_f\rangle$ for $^{239}$Pu will cause the variations of neutron flux density $\phi$ and neutron concentration $n$. This means that an expression for the fission rate of $^{239}$Pu in uranium-plutonium cycle (1), which is analogous to (4), will be more complicated in the multigroup approximation.

However, in spite of this difficult for analytical determination of variation of plutonium fission cross-section, it is possible to show (without loss of generality) the lower estimation of relative variation in the case of multigroup approximation

$$\frac{\Delta\lambda_{Pu}}{\lambda_{Pu}} \sim \frac{\Delta\sigma_f^{Pu}}{\sigma_f^{Pu}} \geq 1 \ . \tag{7}$$

Now let us show to what value of uncertainty of georeactor antineutrino spectrum with oscillations the relative error of plutonium fission rate (5) leads. For this purpose we write down the theoretical form of measured total energy spectrum $dn_i/dE \equiv n_i(E)$ in the $i$th energetic bin

$$n_i(E) = m_\lambda v_i(E) , \tag{8}$$

where

$$m_\lambda = \lambda_{Pu}\Delta t, \quad v_i(E) = \frac{\varepsilon_i N_p}{4\pi L^2}\sum_{j,i}\alpha_i\rho_{ji}(E)\sigma_{vp}(E)p(E,L) , \tag{9}$$

$$p(E,l) = 1 - \sin^2(2\theta_{21})\sin^2\left(\frac{\pi l}{L}\right), \quad \text{where} \quad L(E) = \frac{2.48\,E[MeV]}{\Delta m_{12}^2[eV^2]}[m] . \tag{10}$$

Here $m_\lambda$ is the total number of fissions during the exposure time $\Delta t$ determined by the fission rate $\lambda_{Pu}$; $v_i(E)$ is the average number of detected antineutrino per fission in the $i$th energetic bin; $\varepsilon$ is the detection efficiency for positrons in the inverse $\beta$-decay reaction; $N_p$ is the number of protons in the detector sensitive volume; $\Delta t$ is the exposure time; $p(E, L)$ is the neutrino oscillation probability at the appropriate parameters of mixing and energy $E$ at a distance of $l$ from the reactor; $L$ is the oscillation length; $\theta_{21}$ is the mixing angle; $\Delta m^2_{12} \equiv |m_2^2 - m_1^2|$ is the mass squared difference; $(1/4\pi L^2)$ is the effective solid angle; $\sigma_{vp}$ is the antineutrino-proton interaction cross-section of inverse $\beta$-decay reaction with the corresponding radiation corrections; $\Sigma \alpha_i \rho_i(E)$ is the energy antineutrino spectrum of nuclear fuel in the $i$th energetic bin, MeV/fission; $\alpha_i$ is the part of $i$th isotope.

Here it should be noted that, in general, normalized energy antineutrino spectra corresponding to the different values of reactor heat power it is possible to consider as self-similar. This fact considerably lightens its further analysis. At the same time, a self-similarity takes place only for equilibrium neutrino spectra [42, 43], which is typical for stationary processes in reactor core. And conversely, when processes in the reactor core are nonstationary, a self-similarity of equilibrium neutrino spectra is violated. This means, if, for example, the variations of neutron energy spectrum (and therefore the variations of mass yields induced by the fission of $^{239}Pu$) in the reactor core are considerable, the shapes of corresponding neutrino spectra are not self-similar. Therefore calculated ("stationary", i.e., equilibrium) spectra and corresponding experimental ("nonstationary") neutrino spectra are differ up to 10-15 % and higher [42, 43]. The non-equilibrium effect of neutrino spectra will be considered more specifically in Sec. 6.

Obviously, due to the stochastic change of heat power of nonstationary nuclear georeactor due to the variations of fission cross-section $\langle\sigma_f\rangle$ of the $^{239}Pu$ nuclei and georeactor neutrino spectrum shape (9) the relative uncertainty of georeactor antineutrino spectrum $n_i^{grn}(E)$ with oscillations in the $i$th energy bin (with an allowance for Eqs. (6)-(9)) looks like

$$\frac{\Delta n_i^{grn}}{n_i^{grn}} \simeq \left[\left(\frac{\Delta \lambda_{Pu}}{\lambda_{Pu}}\right)^2 + \left(\frac{\Delta \rho_i}{\rho_i}\right)^2\right]^{1/2} \geq 1 \ , \qquad (11)$$

where $(\Delta\rho_i/\rho_i) \geq 10\%$ is the relative uncertainty due to nonstationarity of georeactor neutrino spectrum shape.

Therefore the lower estimation of uncertainty of total antineutrino spectrum with oscillations with an allowance for Eq.(11) and the contribution of uncertainty (4.14%) of antineutrino spectrum $n_i^{Jap}(E)$ from the Japanese reactors [4] take on form

$$\Delta n_i \geq \left[(0.0414\, n_i^{Jap})^2 + (n_i^{grn})^2\right]^{1/2}. \tag{12}$$

Note that just this uncertainty is shown in Figure 4 as a violet band.

Now we are ready to use our model of uranium-plutonium georeactor [17] for the alternative description of the data produced in new KamLAND experiment [4]. Obviously, that the standard methods of obtaining consistent estimates (e.g., the maximum-likelihood method) normally used for the determination of the oscillation parameters ($\Delta m_{12}^2$, $\sin^2 2\theta_{12}$) [1-5] must take into account one more reactor, or, more specifically, take into account the antineutrino spectrum of georeactor with the power of 30 TW which is located at a depth of $L \sim 5.2 \cdot 10^6$ m. However, following [17], we propose here a simple estimating approach. The results of its application show that the hypothesis of the existence of a georeactor on the boundary of liquid and solid phases of the Earth's core does not conflict with the experimental data.

So, we proceed as in [17] if CPT-invariance is assumed, the probabilities of the $v_e \rightarrow v_e$ and $\tilde{v}_e \rightarrow \tilde{v}_e$ oscillations should be equal at the same values $L/E$. On the other hand, it is known that the variations of $\Delta m^2$ dominate over the more stable small variations of angle $\theta$ at the spectral distortion (oscillations) of "solar" neutrino spectrum. Therefore we can assume (on the grounds of CPT-theorem) that the angle which is determined by the experimental "solar" equality $\tan^2\theta_{12} = 0.447$ [44] may be used as the reference angle of mixing in the KamLAND-experiment.

Finally, following the computational ideology of [17] we give the results of verification of the optimal oscillation parameters ($\Delta m_{21}^2 = 2.5 \cdot 10^{-5}$ eV$^2$, $\tan^2\theta_{12} = 0.437$) by comparing the theoretical (which takes into account the georeactor operation) and experimental spectra of reactor antineutrino based on the KamLAND data over the period of 2002-2007 (Figure 4). We compare also in Figure 5 the $\chi^2$-profiles for our georeactor hypothesis and KamLAND nonzero hypothesis, which does not take into account an uncertainty of reactor antineutrino spectrum (see Section 6).

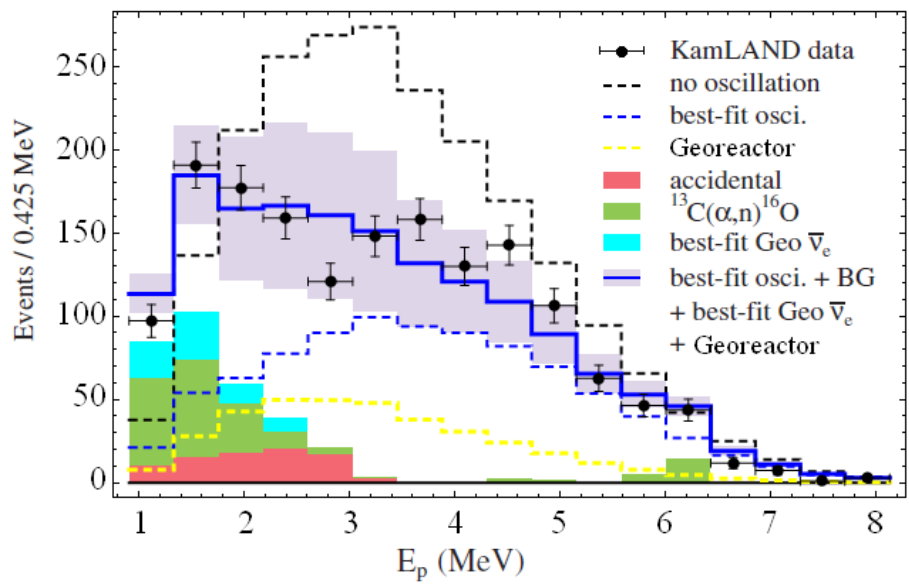

**Figure 4.** Prompt event energy spectrum of $\tilde{\nu}_e$ candidate events (2002-2007). The shaded background and geo-neutrino histograms are cumulative. Statistical uncertainties are shown for the data; the violet band on the blue histogram indicates the event rate systematic uncertainty. The georeactor power is 19.5 TW. The georeactor is at a distance of 5098 km from the KamLAND-detector (see explanation in the text and Table 1).

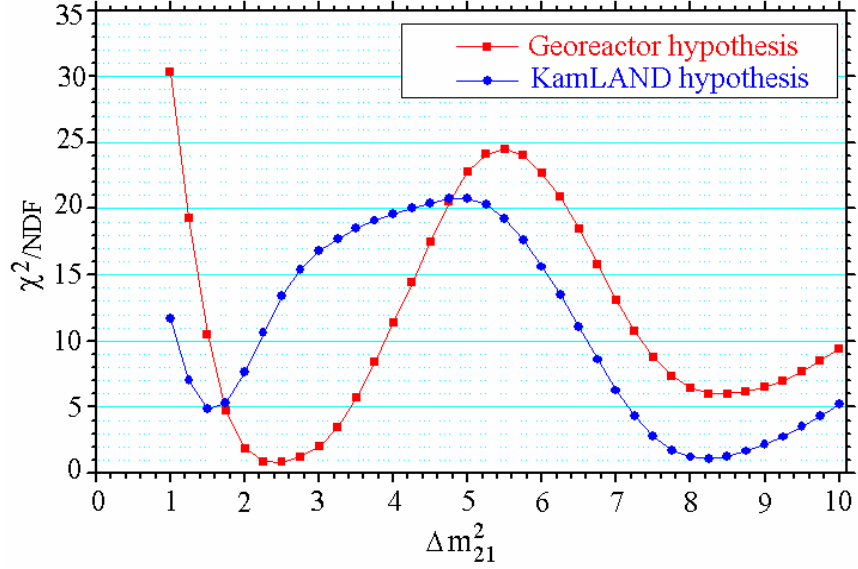

**Fig. 5.** Dependences of $\chi^2$/NDF on the mass squared difference $\Delta m_{21}^2$ corresponding to Kam LAND-hypothesis without georeactor (**blue** line, tg$^2\theta_{12}$=0.56 **[4]**) and our georeactor hypothesis (red line, tg$^2\theta_{12}$=0.437).

In spite of low statistics of neutrino events ($\leq 150$ events/bin), the theoretical reactor antineutrino spectrum (which takes into account a soliton-like nuclear georeactor with the power of 19.5 TW) describes with an acceptable accuracy the experimental KamLAND-data (Figure 4) [4]. Below we pay attention to some important moments.

***Singularities of the low antineutrino statistics accounting***. It is obvious, that a low antineutrino statistics is the reason of high inhomogeneity of filling event density of an antineutrino spectrum (which is continuous by its nature). This inhomogeneity intensifies due to energy discretization of spectrum (0.425 MeV in KamLAND-experiment). As a result the formal procedure of events integration within the one energy bin ($\Delta E$=0.425 MeV) can generate substantial deviations of the number of events (per bin) from its true average value. To observe this effect it would be necessary to decrease considerably the width of energy bin or, that is the same, to decrease the width of detector energy window. However, since it is impossible, we will attempt to show this effect in the following way.

As follows from Figure 4, the significant disagreement between the experimental and theoretical antineutrino spectra is observed for the 5, 7 and 9 bins. Therefore, if the apparent condition

$$P(l)=(1-p(E_5,l)p(E_7,l)p(E_9,l))=\max , \qquad (13)$$

to impose on oscillations of nuclear georeactor antineutrino spectrum, then by Eq. (10) and the average energies of bins $E_5$=2.8 MeV, $E_7$=3.7 MeV and $E_9$=4.5 MeV (see Figure 4) we can obtain a series of the values $l$ for distances from KamLAND-detector to the supposed location of georeactor on the surface of Earth's solid core (Figure 6).

$$l=5365, \quad 5968, \quad 6400, \quad 6830, \quad 7410 \quad km . \qquad (14)$$

Now we return to the problem of low antineutrino statistics. Fulfilment of condition (13) for given distances (14) makes it possible to recalculate a georeactor antineutrino spectrum (Figure 4) for these distances by Eqs. (8)-(10). Proceeding from a low antineutrino statistics (in energy bins $E_5$=2.8 MeV, $E_7$=3.7 MeV и $E_9$=4.5 MeV), the following variants of the location of a georeactor on the Earth solid core surface are most acceptable: (i) a georeactor with the thermal power of 30.7 TW at 6400 km distance from KamLAND detector (Figure 7a); (ii) a georeactor with the thermal power of 34.7 TW at 6830 km distance from KamLAND detector (Figure 7b); georeactors of equal thermal power but with the total power of 32.6 TW at 6400 and 6830 km distance from KamLAND detector (Figure 7c).

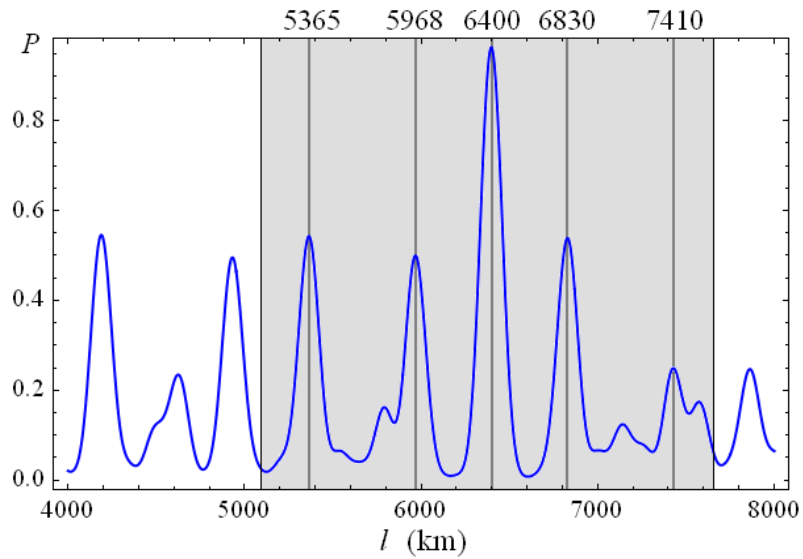

**Figure 6.** The spatial frequency distribution *P(l)* of oscillating georeactor antineutrinos with energies $E_5$=2.8 MeV, $E_7$=3.7 MeV и $E_9$=4.5 MeV. Shaded area corresponds to the continuous series of distances $l$ from the KamLAND-detector to the Earth solid core surface .

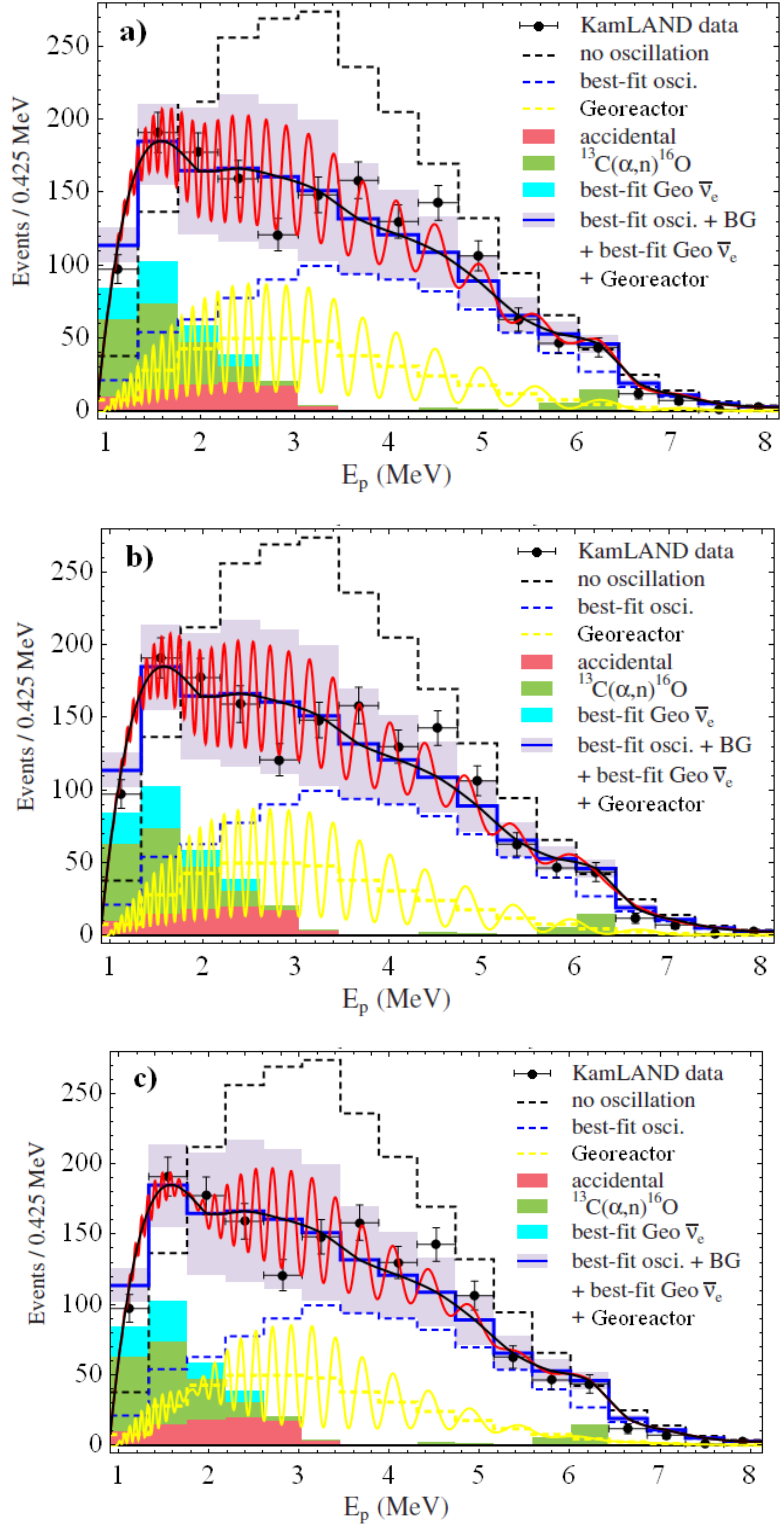

**Figure 7.** The theoretical antineutrino spectrum (blue histogram), which takes into account the nuclear georeactor a) with thermal power of 30.7 TW (yellow histogram) situated at a distance of 6400 km from the KamLAND-detector, b) with thermal power of 34.7 TW (yellow histogram) situated at a distance of 6830 km from the KamLAND-detector, c) with general thermal power 32.6 TW (yellow histogram) situated simultaneously at the distances of 6400 and 6830 km from the KamLAND-detector. In all figures one can see how discrete antineutrino spectra of KamLAND-experiment (blue histogram) and georeactor (yellow histogram) mask the low statistics effect in corresponding continuous antineutrino spectra (red and yellow oscillations).

These results constrain us to recalculate the georeactor thermal power obtained for KamLAND data over the period of 2002-2004 [17]. The parameters of nuclear georeactors obtained by two calculation methods of antineutrino spectra over the periods of 2002-2004 and 2002-2007 are collected in Table 1.

The considered singularities of low antineutrino statistics make it possible not only to determine the possible distances from KamLAND detector to supposed nuclear georeactor on the Earth solid core surface (Figure 6), but to construct the map of located on the Earth surface lines radially conjugate to lines-circumferences formed by the bases of cones with a vertex in KamLAND and generating sides, whose lengths are equal to a corresponding distances from KamLAND detector to the Earth solid core surface (Figure 8).

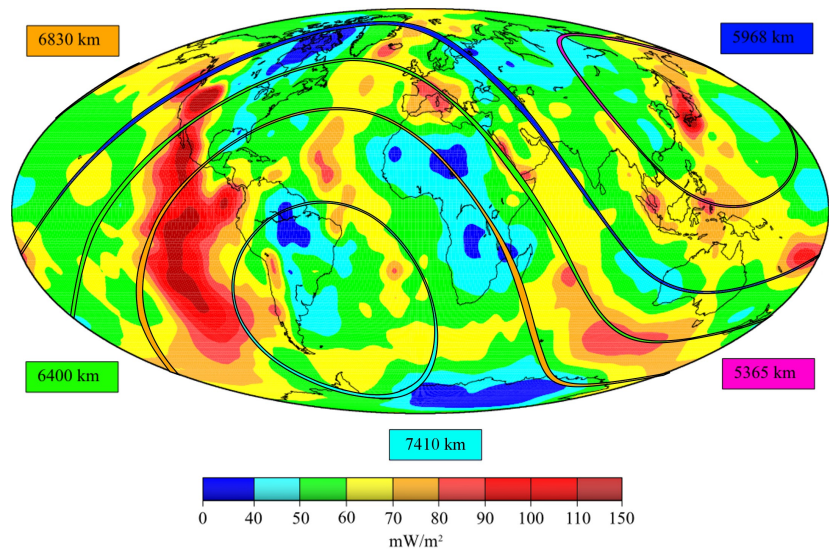

**Figure 8.** Distribution of geothermal power density on the Earth [45] superposed with the conjugated "pseudoreactor" circumferences, which are generated by "georeactor" circumferences located on the perimeters of the bases of cones with a vertex in KamLAND (36.43°N and 137.31°E) and generating sides, whose lengths from KamLAND-detector to the Earth solid core surface are equal to 7410 (sky blue), 6830 (orange), 6400 (green), 5968 (blue) and 5365 (pink) km.

*Non-stationary nature of soliton-like nuclear georeactor*. Analyzing Table 1, we can see that within the framework of modified method for calculation of antineutrino spectrum (Figure 7) the average thermal power of georeactor changes from ~ 50 TW (during the exposure of 749.1 days in 2005 [3], Figure 2) to ~30 TW (at total exposure of 1890.25 days in 2008 [4]). This, undoubtedly, is the reflection of non-stationary nature of georeactor. Taking into account that total exposure (1890.25 days) is sum of two consecutive exposures (749.1 and 1141.15 days, respectively), in fact the change the average thermal power of georeactor is still more, i.e., from ~50 TW over the first exposure to ~20 TW over the second consecutive exposure (see Table 1).

From the physical standpoint, the decrease almost in two times of georeactor thermal power (during the two successive exposures) means that the variances of fission cross-section $\langle\sigma_f\rangle$ for the $^{239}_{94}Pu$ nuclei during total exposure over the period of 2002-2007 also change in two times or, more exactly, go almost 100 % down. It means that the variance of fission cross-section $\langle\sigma_f\rangle$ for the $^{239}_{94}Pu$ nuclei reaches practically 100% and this is in good agreements with our estimation of variance of fission rate (7).

**Table 1.** Nuclear georeactor thermal power depending on a distance to detector and exposure time in the KamLAND and Borexino experiments

| Period | 2002-2004 | 2002-2007 | 2005-2007 | 2002-2009 | 2008-2009 | 2008-2009 |
|---|---|---|---|---|---|---|
| Experiment | KamLAND | | | | | Borexino |
| Exposure, days | 749.10 | 1890.25 | 1141.15 | 2135 | 244.75 | 537.20 |
| Distance, km | Nuclear Georeactor power, TW | | | | | |
| 5098 | 30.0 | 19.5 | 12.6 | 17.3 | 4.7 | - |
| 6400 | 47.3 | 30.7 | 19.8 | 28.0 | 7.1 | - |
| 6830 | 53.4 | 34.7 | 22.4 | 31.6 | 7.7 | - |
| 6400+6830 | 50.2 | 32.6 | 21.1 | 29.7 | 7.3 | - |
| 6711 | - | - | - | | | 5.0 |

At the same time, we asserted before that the main cause of change of fission cross-section $\langle\sigma_f\rangle$ for the $^{239}_{94}Pu$ nuclei in extreme thermodynamics conditions are the temperature variations of fuel medium. Therefore there is a natural question, that (except the georeactor) is reason of the temperature variations of fuel medium or, more exactly, what is physical nature of independent source of the temperature variations of fuel medium, which in the end plays the role of external modulator of nuclear georeactor thermal power. Answer on this very important question related to finding out of physical reasons of non-stationary nature of soliton-like nuclear georeactor will be given in the second part of this paper [46].

Briefly summing the results of this section, we can say that in spite of the low statistics of neutrino events (≤ 150 events/bin), the theoretical reactor antineutrino spectrum (which takes into account the soliton-like nuclear georeactor with the power of 30 TW) describes with acceptable accuracy the experimental KamLAND-data [4] (see Figures 4 and 7). Here we pay attention to some important moments. Firstly, the average georeactor heat power is changed from ~50 TW at the exposure time of 749.1 days in 2005 [3] (Figure 2) to ~ 30 TW at total exposure of 1890 days in 2008 [4] (Figure3). This reflects the nonstationary nature of the georeactor.

## 4. The Borexino and KamLAND experiments and triangulation of soliton-like nuclear georeactors location

As is generally known, the first stage of Borexino experiment (Laboratory Nationali del Gran Sasso, Italy) [47] was recently completed, ideology of neutrino measurements in which is practically analogical the neutrino measurements in KamLAND experiment. It means that the joint use of the Borexino and KamLAND data opens up non-trivial possibility for the solving of very important problem of spatial identification of nuclear georeactor location on the Earth solid core surface or, otherwise speaking, the triangulation of the soliton-like nuclear georeactor location on the boundary of the liquid and solid phases of the Earth core.

***Borexino antineutrino spectrum (exposure over the period of 2008-2009).*** We give here the alternative analysis of the Borexino data collected between December 2007 and December 2009, corresponding to 537.2 days of live time [47]. The fiducial exposure after cuts is 252.6 ton· yr. The determination of the expected neutrino signal from reactors, which, as usual, was calculated by Eq. (8), required the collection of the detailed information on the time profiles of power and nuclear fuel composition for nearby reactors. In Eq. (8) the main contribution comes from 194 reactors in Europe, while other 245 reactors around the world contribute only 2.5% of the total reactor signal. Information on the nominal thermal power and monthly load factor for each European reactor originate from IAEA and EDF [47].

It is important to note that to describe the antineutrino specta in the Borexino experiment the parameters of mixing ($\Delta m^2_{21}$=7.65· $10^{-5}$ eV$^2$, $\sin^2\theta_{12}$=0.304 [48]) based on the global three-flavour analysis of solar (SNO) and reactor (KamLAND) experimental data were used. At the same time, in alternative describing the the Borexino data (Figure 9) we use the parameters of mixing $\Delta m^2_{21}$ =2.5· $10^{-5}$ eV$^2$, $\tan^2\theta_{12}$ = 0.437 or, in other words, $\sin^2\theta_{12}$ = 0.304 obtained by our model, which takes into account the existence of natural nuclear reactor on the boundary of the liquid and solid phases of the Earth core [17].

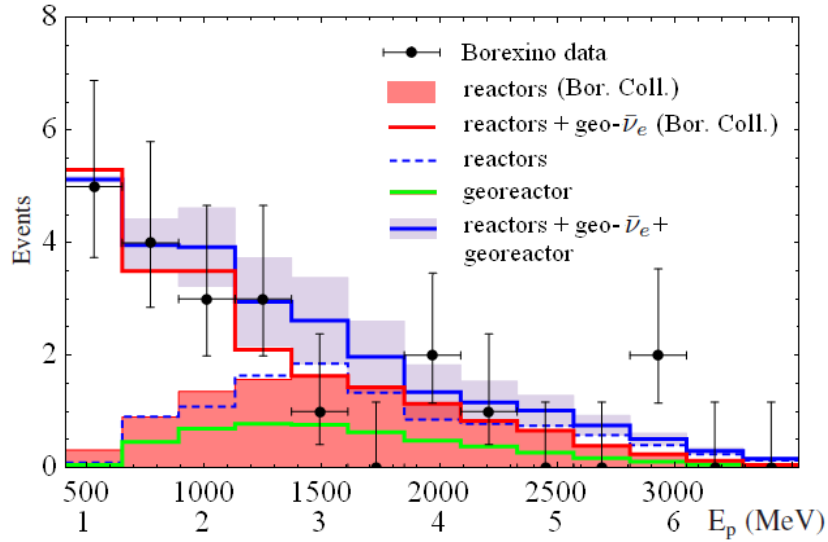

**Figure 9.** The $\tilde{\nu}_e$ energy spectra in Borexino [47]. Experimental points (solid black dots with error bars) together the total expectation obtained in Borexino experiment (red solid histogram) [3] and in the presented paper (blue solid histogram). Also shown are the expected neutrino spectrum from Europe's reactors calculated by our parameters of mixing (blue dashed histogram) and obtained in Borexino-experiment (red shaded area). The expected neutrino spectrum from the georeactor of 5 TW located at a distance of 6711 km from the Borexino-detector (green line) is also shown (see explanation in the text).

It is obvious, that in spite of very low statistics of neutrino (reactor) events (1-2 events/bin) the theoretical reactor antineutrino spectrum (which takes into account a soliton-like nuclear georeactor with the power of 5 TW) describes with an acceptable accuracy the experimental Borexino data (Figure 9) [47]. Note that the lower estimation of uncertainty of total antineutrino spectrum with oscillations (which is headlined in violet colour in Fig 9) was calculated by Eq. (12) at the uncertainty of antineutrino spectrum $n_i^{Euro}(E)$ from European reactors equal to 5.38% [47].

***Singularities of the low antineutrino statistics accounting***. As follows from Figure 9, the considerable disagreement between the experimental and theoretical antineutrino spectra is observed for 5, 6, 7 and 11 bins. Therefore, if the apparent condition

$$P(l)=(1-p(E_5,l))(1-p(E_6,l)p(E_7,l)p(E_{11},l))=\max , \qquad (15)$$

to impose on oscillations of nuclear georeactor antineutrino spectrum, then by Eq. (15) for $p(E,l)$ and the bin average energies $E_5$, $E_6$, $E_7$ and $E_{11}$ (see Figure 9) we can obtain a series of the values $l$ for possible distances from Borexino-detector to the supposed location of georeactor on the surface of Earth's solid core (Figure 10).

$$l=5310, \quad 5400, \quad 6310, \quad 6711, \quad 7128, \quad 7490 \quad km . \qquad (16)$$

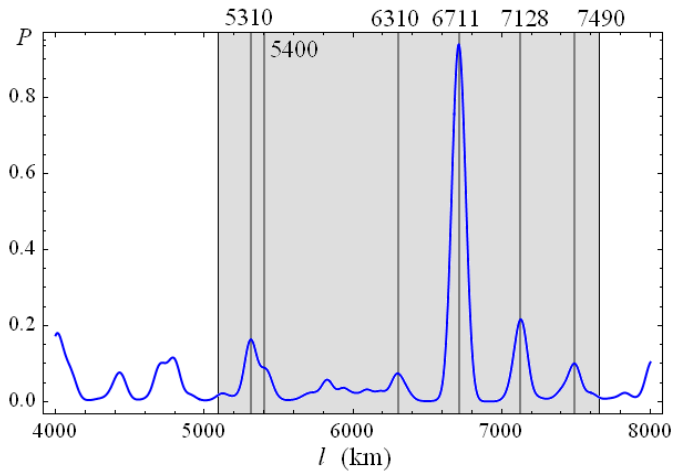

**Figure 10.** The spatial frequency distribution *P(l)* of oscillating georeactor antineutrinos with energies corresponding to 5, 6, 7 and 11 bins in Borexino-spectrum in Figure 9. Shaded area corresponds the continuous series of the distances *l* from the Borexino-detector to the surface of the Earth solid core.

Returning to the problem of low antineutrino statistics, note that the fulfilment of the condition (15) for given distances (16) makes it possible to recalculate by Eqs. (8)-(10) a georeactor antineutrino spectrum (Figure 9) for these distances. Proceeding from a low antineutrino statistics (in energy bins $E_5$, $E_6$, $E_7$ and $E_{11}$), the most acceptable location of a georeactor on the Earth solid core surface is at 6711 km distance from the Borexino detector (Figure 10). It is easy to show that the thermal power of such a nuclear georeactor is ~5 TW.

It is necessary to note that the restriction on the nuclear georeactor thermal power obtained within the frameworks of the Borexino experiment is ~3 TW [47]. Though this restriction is obtained within the framework of nonzero georeactor hypothesis, it does not take into account the high uncertainty of georeactor antineutrino spectrum. The consequences of neglect of this uncertainty come into a question in section 6.

***Triangulation of the locations of soliton-like nuclear georeactors.*** By triangulation of the KamLAND and Borexino data we have constructed the "pseudogeoreactor" coordinate location conjugate to the real location of soliton-like nuclear georeactors on the boundary of the liquid and solid phases of the Earth core (Figure 11).

Analyzing Figures 6, 7 and also Figure 10 (by which it is possible to determine the most probable distances between the detector and a nuclear georector), we have divided georeactors into two groups - operating reactors and probable low-power reactors (Figure 11).

Naturally, a question is the following: What is the cause of sufficiently high degree of correlation between "conjugate pseudoreactors" and the regions of higher geothermal power of the Earth in Figure 11? Below we will consider physical reasons causing such a correlation.

Here it should be reminded, that according to our assumption [17], nuclear georeactors are located in the thin uranium-containing high-density layer (about 2.2 km) [33], which is the peculiar physical boundary of the liquid and solid phases of the Earth core. According to the results of seismic tomography [33], this layer has a mosaic structure, whose typical size is ~ 200 km. This means that the spatial history of nuclear burning wave or, in other words, "burning spot" is completely determined by the area of one patch of a mosaic structure. Since a figure on the Earth surface conjugate to the single patch of a mosaic structure has the typical size ~ 1000 km, this value will be the size of domain of uncertainty for the "conjugate burning spot" on the Earth surface.

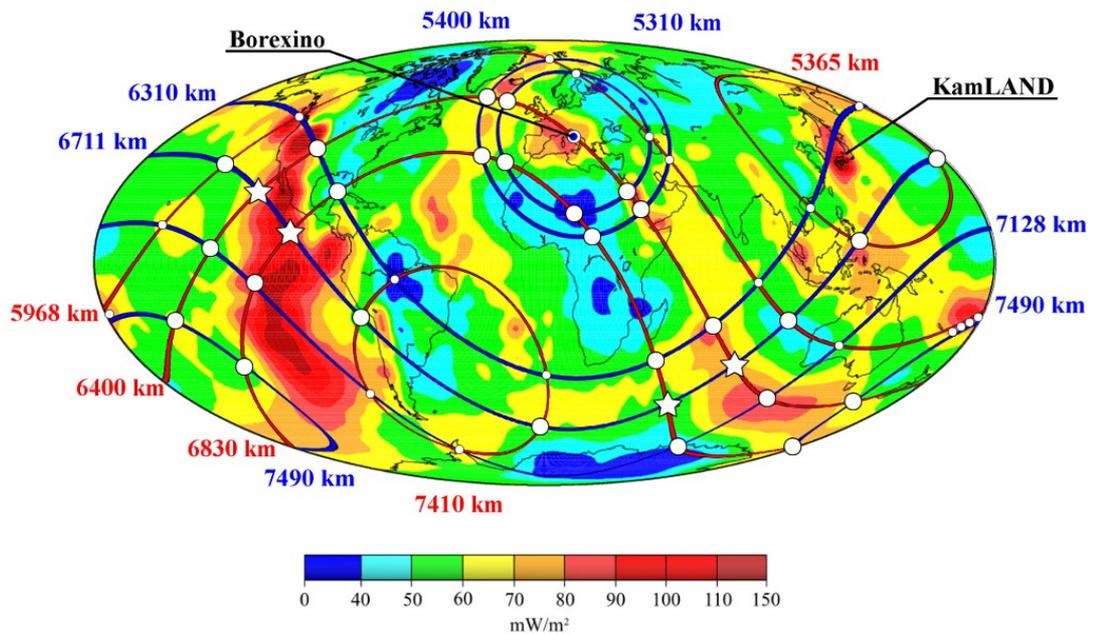

**Figure 11.** Distribution of geothermal power density on the Earth [45] superposed with the conjugate "pseudogeoreactor" ellipsoidal closed curves, which were built on basis of KamLAND (red lines) and Borexino (blue lines) experimental data. (☆) – operating nuclear georeactors; (O) and (∘) – nuclear georeactors, whose power (if they are operating) is an order of magnitude and more less than the thermal power of reactors designated by (☆).

Here it should be reminded, that according to our assumption [17], nuclear georeactors are located in the thin uranium-containing high-density layer (about 2.2 km) [33], which is the peculiar physical boundary of the liquid and solid phases of the Earth core. According to the results of seismic tomography [33], this layer has a mosaic structure, whose typical size is ~ 200 km. This means that the spatial history of nuclear burning wave or, in other words, "burning spot" is completely determined by the area of one patch of a mosaic structure. Since a figure on the Earth surface conjugate to the single patch of a mosaic structure has the typical size ~ 1000 km, this value will be the size of domain of uncertainty for the "conjugate burning spot" on the Earth surface.

On the other hand, we know that the time of heat transmission from the "burning spot" to the "conjugate burning spot" is of order $10^9$ years. Taking into account the average velocity of nuclear burning wave (~ 1 m/year), we obtain that for $10^9$ years the "burning spot" will cover the total distance of $10^6$ km on the Earth surface. Because this distance is easily go in the area of one patch of a mosaic structure in the form of a certain fractal-broken curve, the domain of uncertainty of thermal flow (on the Earth surface) coincides with the domain of uncertainty of the "conjugate burning spot". Exactly this coincidence is the reason of good correlation between "conjugate burning spots" location and the ranges of higher geothermal power in Figure 11. In other words, a map of geothermal power distribution on the Earth is simultaneously the rough approximation of acting or before acting nuclear georeactors, whose location is determined in this case with an accuracy up to 1000 km.

It is interesting, that if in places, where the nuclear georectors location is supposed, any considerable geothermal heat release is absent (see Figure 11), large so-called solitary volcanoes (hot spots), for example, in the Central and South-East Africa (Figure 12) or active volcanoes, for example, Erebus in the Antarctic (Ross Sea coast), Deception (South Scotch Islands) and discovered recently nameless volcano (Hudson Mountains in the West Antarctic [49]) are necessarily there.

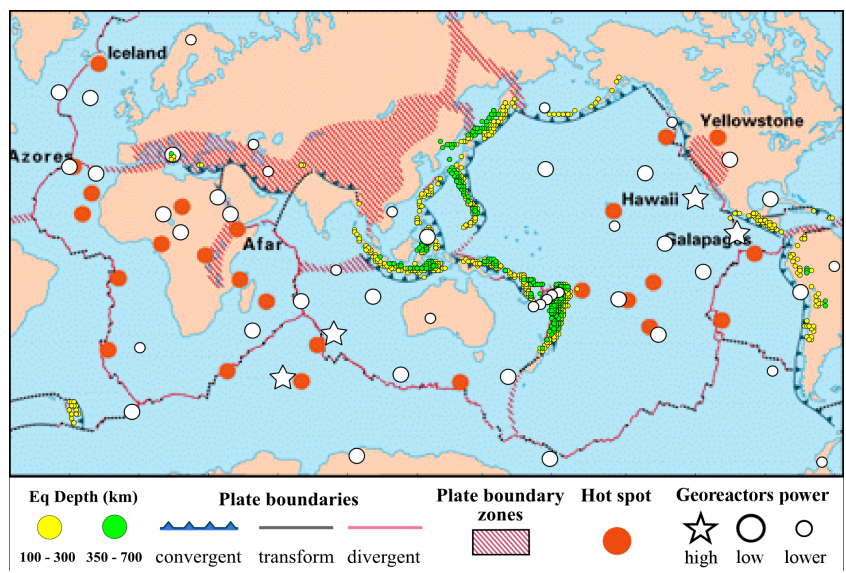

**Figure 12.** The map of spatial location of "pseudureactors", hot spots and deep-focus earthquakes over the period of 1993-2006. The map of "pseudureactors" and earthquakes was built on basis of the plate boundaries map [50].

## 5. Non-stationary soliton-like nuclear georeactor and
## new KamLAND antineutrino spectrum (exposure over the period of 2002-2009)

When this paper was written the immediate data of KamLAND experiment obtained over the years of 2002-2009 was published [5]. In spite of this we decided not to change the structure of the paper but to add the analysis of these data to existing material, because such an analysis is the natural illustration of inner consistency of the considered above georeactor hypothesis.

***KamLAND antineutrino spectrum.*** We give here the alternative description of KamLAND-data [5] collected from March 9, 2002, to November, 4, 2009, corresponding to 2135 days of live time. The number of target protons within the 6.0-m-radius spherical fiducial volume is calculated to be $(5.98 \pm 0.12) \cdot 10^{31}$ for the combined data set, which corresponds to an exposure to electron antineutrino $\tilde{\nu}_e$ of $3.49 \cdot 10^{32}$ proton-years. The determination of the expected neutrino signal from reactors, which was traditionally calculated by Eq. (8), required the collection of the detailed information on the time profiles of power and nuclear fuel composition for nearby reactors. The relative fission yields, averaged over the entire live-time period, for isotopes ($^{235}$U:$^{238}$U:$^{239}$Pu:$^{241}$Pu) are (0.571:0.078:0.295:0.065), respectively. In Eq. (8) the main contribution comes from 56 Japanese nuclear power reactors, while the contributions from Korean reactors and the remainder of the global nuclear power industry is estimated to be $(3.4 \pm 0.3)\%$ and $(1.0 \pm 0.5)\%$ of the total reactor signal, respectively. Information on the nominal thermal power and monthly load factor for each Japanese reactor originate from consortium of Japanese electric power companies [5].

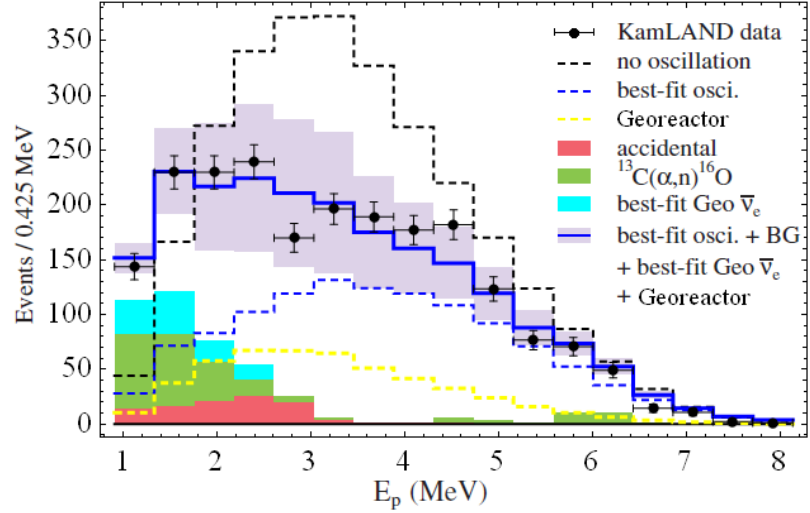

**Figure 13.** Prompt event energy spectrum of $\tilde{\nu}_e$ candidate events (exposure over the years of 2002-2009). The shaded background and geo-neutrino histograms are cumulative. Statistical uncertainties are shown for the data; the violet band on the blue histogram indicates the event rate systematic uncertainty in the framework of georeactor hypothesis. The total georeactor power is 29.7 TW. Georeactors are located at a distance of 6400 and 6830 km from the KamLAND-detector (see explanation in the text and Table 1).

It is obvious, that the theoretical reactor antineutrino spectrum (which takes into account a soliton-like nuclear georeactor with the power of 29.7 TW (see Table 1)) describes with an acceptable accuracy the new experimental KamLAND-data (Figure 13). In so doing the lower estimation of uncertainty of total antineutrino spectrum with oscillations is calculated by Eq. (12) with an allowance for the contribution of the uncertainty 4.5 % (which corresponds to the variant DS-2 [5]) of total antineutrino spectrum from the Japanese reactors.

From Table 1 it follows that the average thermal power $W_t$ of nuclear georeactor sharply decreases in KamLAND experiments corresponding to the exposures over the periods of 2002-2004, 2002-2007 and 2002-2009. Such a power jump indicates that the nuclear georeactor is strongly nonstationary. It is very important fact for the correct calculation of reactor geoneutrinos, which in the end are the integral part of KamLAND antineutrino spectrum (within the framework of georeactor hypothesis). To illustrate such a strong nonstationarity we give below the change of the georeactor average thermal power over the period of 2002-2009.

Using the average values of nuclear georeactor thermal power $W_t$ reconstructed within the framework of georeactor hypothesis (Table 1), which correspond to exposures over the years of 2002-2004, 2002-2007 and 2002-2009, it is possible to determine the values $W_t$ corresponding to "latent" exposures over the years of 2005-2007 and 2008-2009 by obvious expression

$$W_t = \frac{t_1}{t_1+t_2} W_{t_1} + \frac{t_2}{t_1+t_2} W_{t_2}, \quad \text{where} \quad t = t_1 + t_2 . \qquad (17)$$

The values of nuclear georeactor thermal power $W_t$ extended in that way with consideration of exposure over the period of 2002-2004 (see Table 1) make it possible to build the time evolution of the georeactor average thermal power $W$ over the years of 2002-2009 (Figure 14).

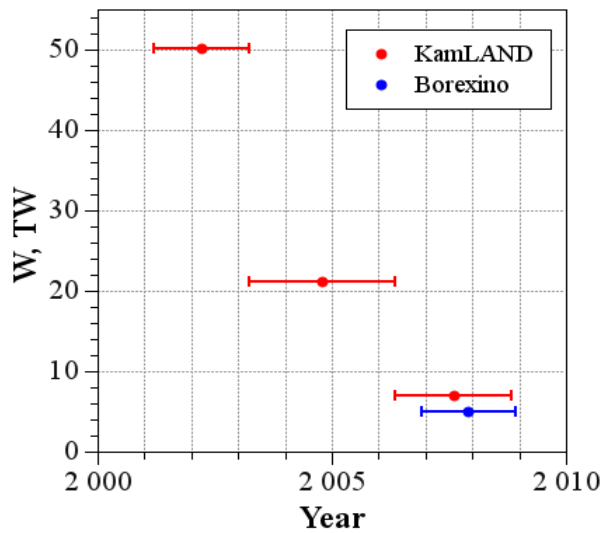

**Figure 14.** Evolution of reconstructed average thermal power $W$ of nuclear georeactor over the years of 2002-2009.

Thus, the sharp change of the georeactor average thermal power $W$ during the years of 2002-2009 must be necessarily taken into account in the calculation of reactor geoneutrino spectra, which within the framework of georeactor hypothesis are component of KamLAND antineutrino spectra. For that within the framework of traditional calculations of effective neutrino spectra of both individual nuclides and their mixture [42, 43] it is necessary to take into account the high degree of nonequilibrium of neutrino spectra due to strong nonstationarity of nuclear georeactor operating [42, 43].

It is very important to note, that in the real experiment exactly the high degree of nonequilibrium of neutrino spectra due to the "latent" unstationarity of radiation source can become the reason of sharp change of the expected "equilibrium" shape of resulting neutrino spectrum. In this sense, ignoring of the high degree of neutrino spectrum nonequilibrium or, in other words, description of experimental effective neutrino spectrum by the equilibrium neutrino spectra of individual nuclides or their mixture can result in serious mistakes in fitting the experimental neutrino spectrum. Let consider this more detailed.

## 6. On some important features of alternative treatment procedure of KamLAND experimental data

*Time variation of the reactor antineutrino flux and upper limit of power georeactor.* In the paper of KamLAND-collaboration [2], where the results of the second exposure (551.1 days) are analyzed, the original and very interesting method for determination of antineutrino rate suppression factor, which describes the degree of distortion of antineutrino spectrum, is presented. With that end in view the time variation of the reactor antineutrino flux assuming no antineutrino oscillation is estimated (see Fig. 15a).

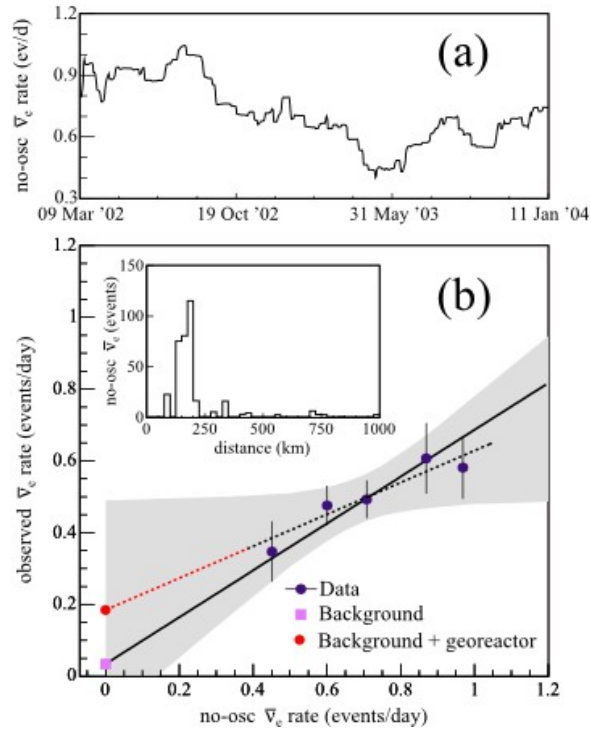

**Fig.15.** Adapted from [2]. (a) Estimated time variation of the reactor $\tilde{\nu}_e$ flux assuming no antineutrino oscillation. (b) Observed $\tilde{\nu}_e$ event rate versus no-oscillation reactor $\tilde{\nu}_e$ flux. Data points correspond to intervals of approximately equal $\tilde{\nu}_e$ flux. The dotted black line is a fit [2]; the 90% C.L. is shown in gray [2]. The solid black line is a fit constrained to the expected background [2]. The dotted red line is continuation of the dashed black line, whose intercept in this case is equal to the sum of expected background and $\tilde{\nu}_e$ flux from unknown source (for example, a georeactor). The reactor distance distribution for $\tilde{\nu}_e$ events in the absence of oscillation is shown in the inset.

Analysis of the linear dependence of the observed $\tilde{\nu}_e$ event rate on the no-oscillation reactor $\tilde{\nu}_e$ flux and assumption that the slope can be interpreted as $\tilde{\nu}_e$ rate suppression factor and the intercept as the reactor-independent constant background rate (Fig. 15b) are the main points of this method. The solid black line in the Fig. 15b is the linear KamLAND-fit (90% C.L.) constrained to the expected background [2]. As analysis of this experiment shows, the value of this expected background corresponds to the sum of background contributions from different background nuclear reactions to the $\tilde{\nu}_e$ signal above 2.6 MeV and is equal to 17.8± 7.3 events. Here it is important to note that to obtain such a fit, the authors of Ref. [2] use the tough assumption that "…the intercept is consistent with known background, but substantially larger backgrounds cannot be excluded; hence this fit does not usefully constrain speculative sources of antineutrinos such as a nuclear reactor at the Earth's core [15]…". In other words, being applied in analysing the observed $\tilde{\nu}_e$-spectra, this assumption is equivalent to acceptance of the so-called zero KamLAND-hypothesis, which a priori eliminates the existence of nuclear georeactor or, at least, rejects the significant influence of additional $\tilde{\nu}_e$ flux from the nuclear georeactor (even if it exists) on the observed $\tilde{\nu}_e$-spectra. In the last case, taking into account the parameters of mixing obtained within the framework of zero KamLAND-hypothesis, the predicted KamLAND rate for typical 3 TW georeactor scenarios is comparable to the expected 17.8± 7.3 event background and would have minimal impact on the analysis of the reactor power dependence signal [2].

On the other hand, the simple analysis of the dashed black line in Fig. 15b shows that the alternative, i.e., nonzero, KamLAND-hypothesis, which concedes the existence of nuclear georeactor with considerable heat power, has full rights to life. Note that the dotted black line is a fit [2], the 90% C.L. is shown in gray [2] and the dotted red line is extended by us the dashed black line, whose intercept in this case is equal to the sum of expected background and $\tilde{\nu}_e$ flux from unknown source.

Below we assume that the nuclear georeactor plays the role of unknown $\tilde{\nu}_e$ source. fission product. Let us estimate its heat power $W$. It will be recalled that for determination of reactor power in neutrino experiments, according to Eqs, (8)-(10) and other things being equal it is necessary to know (i) location of the georeactor, i.e., the distance between the georeactor and detector, (ii) fuel composition and correspoding equilibrium (or nonequilibrium[*]) concentrations of fission products, (iii) georeactor antineutrino spectrum uncertainty, (iiii) the survival probability $p$ or the neutrino oscillation parameters. We consider that for the nuclear georeactors with the heat power $W_0$ ~3 TW (the zero KamLAND-hypothesis [2] and $W$ (the alternative nonzero KamLAND-hypothesis) the conditions (i) and (ii) are identical. Then, taking into account that the intersepts and slopes of stright lines in Fig. 15b corresponding to the solid black line (the zero KamLAND-hypothesis) and the dotted red line (the alternative nonzero KamLAND-hypothesis) are $^{0}n_{v}$ ~0.32, $^{\Sigma}n_{v}$ ~0.195 and $p_0$~0.6, $p$~0.4, respectively, it is easy to obtain the approximate estimation of the nuclear georeactor heat power $W$ within the framework of alternative nonzero KamLAND-hypothesis:

$$\frac{^{\Sigma}n_{v}}{^{0}n_{v}} = \frac{pW + p_0 W_0}{p_0 W_0} \quad \rightarrow \quad W \sim 22 \quad TW \quad . \tag{18}$$

---
[*] The features in the calculations of nonequilibrium neutrino spectra see below (section 6).

It is necessary to notice that the survival probability $p$~0.4 does not characterized now by the neutrino oscillation parameters computed for the zero hypothesis.

In our opinion the given example, which shows some the refinements of application of the time variation of reactor antineutrino flux, is extremely obvious and significant since it substantiates in a natural way the possibility of existence of the two alternative (but peer from the physical standpoint) hypotheses for the interpretation of experimental KamLAND-data.

In this sense it would be interesting to consider the results of Fogli et al. [51], who have analyzed the KamLAND neutrino spectra in energy and time exactly for the second exposure [2]. They acted on the premise that while the energy spectrum KamLAND events allows the determination of the neutrino oscillation parameters, the time spectrum can be used to monitor known and unknown neutrino sources. By using available monthly-binned data on event-by-event energies in KamLAND and on reactor powers in Japan, they performed a likelihood analysis of the neutrino event spectra in energy and time, and not only confirmed the results of KamLAND-collaboration, but within the framework of nonzero hypothesis set the upper bound on hypothetical georeactor power ($W \leq 13$ TW at 95% C.L.).

Here a natural question arises:" Why do we obtain the different values of the neutrino oscillation parameters and upper bound on hypothetical georeactor power in comparison with Fogli et al. analysing the same KamLAND-experiment within the framework of the same nonzero hypothesis?" It is obvious that the main reason of this problem is no the ideology of likelihood analysis, whose details are expounded in [51], but absolutely different understanding of physical properties of hypothetical nuclear georeactor, some of which are discussed above (see derivation of Eq. (12)). Let us show this.

In general, the KamLAND unbinned likelihood function £ can be written as [1, 2, 51, 52]

$$\mathcal{L} = \mathcal{L}_{rate} \times \mathcal{L}_{shape} \times \mathcal{L}_{syst}, \qquad (19)$$

where the three factors embed information on the total event rate, on the spectrum shape and on the systematic uncertainties.

According to [51], we remain unchanged the second and third likelihood factor in Eq.19 and consider the first likelihood factor, which can be written as (see also [1, 2, 51, 52]):

$$\mathcal{L}_{rate} = \frac{1}{\sqrt{2\pi}\,\sigma_{rate}} \exp\left[-\frac{1}{2}\left(\frac{N_{theor}(\delta m^2, \sin^2\theta_{12}; \alpha, \alpha', \alpha'') - N_{obs}}{\sigma_{rate}}\right)^2\right], \qquad (20)$$

where ($\delta m^2$, $\sin^2\theta_{12}$) are the mass-mixing parameters, $\alpha$ is the systematic energy offset, $\alpha'$ and $\alpha''$ are free (positive) parameters, $N_{obs}$ is the total number of observed events and the total error is the sum of the statistical and systematic uncertainties

$$\sigma_{rate}^2 = N_{theor} + (sN_{theor})^2, \qquad (21)$$

where $s$ is the part of systematic uncertainty.

Here is key moment which reveals physial distinction between our nonzero hypothesis and nonzero hypothesis by Fogli et al. [51]. In general case, when within the framework of the nonzero hypothesis the additional antineutrino source (i.e., a nuclear georactor) is taken into account, it is necessary also to take into account the uncertainty of georeactor antineutrino spectrum. As shown above (see Eq. (12)), this uncertainty appears due to the change of fission cross-section of $^{239}$Pu (which is the main component of nuclear fuel) with change of nuclear fuel temperature and, in particular, with change of temperature near the Earth's solid core safety (see Fig. 3). As a result, in the case of nonzero hypothesis Eq. (21) must have, acording to Eq. (12), the following form:

$$\sigma^2_{rate} = N_{theor} + \left[s(N_{theor} - N_{grn})\right]^2 + (s_0 N_{grn})^2, \quad s_0 \sim 1 \gg s, \qquad (22)$$

where $N_{theor} = N_{jap} + N_{grn}$, $N_{Jap}$ is the total number of events from Japanenese nuclear reactor, $N_{grn}$ is the total number of events from nuclear georeactor, $s_0$ is the part of systematic uncertainty of the number of georeactor antineutrino.

It is obvious that Eq. (22) in contrast to Eq. (22) admits the high value of the nuclear georeactor heat power. This, in its turn, leads to change of survival probability and, conesquently, to change of the neutrino oscillation parameters. In this sense, it is clear that even very accurate taking account of the time variation of the reactor antineutrino flux (for example, monthly or even daily neutrino flux) and another no less important features of antineutrino spectrum does not lead to considerable change of the antineutrino survival probability (see Fig. 15b, solid black line), if the peculiar uncertainty of georeactor antineutrino spectrum will not be taken into account. And conversely, taking in account such a feature of georeactor antineutrino spectrum, we have obtained new restrictions on the georeactor heat power and corresponding values of the neutrino oscillation parameters (Fig. 5) by likelihood analysis of the KamLAND energy spectrum (Fig. 4) and minimization of corresponding $\chi^2$-function based on Eq. (19).

Returning to the known KamLAND estimation of georeactor heat power, we would like to cite Ref. [4]: "The KamLAND-data, together with solar $\nu$ data, set an upper limit of 6.2 TW (90% C.L.) for a $\tilde{\nu}_e$ reactor source at the Earth's center [15] assuming that the reactor produced a spectrum identical to that of a slow neutron artificial reactor". Although this does not evidently follow from the paper, we suppose that within the framework of likelihood analysis of the KamLAND neutrino spectra in energy and time the authors used the nonzero georeactor hypothesis by adding a 57th reactor at $L$=6400 km to the 56 Japanese nuclear power reactors. At the same time, nontrivial properties of some components of nuclear fuel (for example, the $^{239}$Pu fission cross-section (see Fig. 3)) in this paper as well as in the all another works of KamLAND-collaboration, was not taken into account at all. As shown above, such a neglect of anomalous behavior of the $^{239}$Pu fission cross-section at high temperatures (in the 2500 to 6000 K range (Fig. 3)) implies, according to Eq. (22), the ignoring of high uncertainty of georeactor antineutrino spectrum, which within the framework of maximum likelihood analysis will immediately cause the distortion of "true" values of reactor heat power and corresponding values of the neutrino oscillation parameters.

At last note that nonstationary regime of nuclear georeactor operating caused by change of the $^{239}$Pu fission rate (Eq. (5)) mainly due to the strong variation of the $^{239}$Pu fission cross-section (Eq.7), which is the nonlinear function of medium temperature (see Fig. 3), is the main reason of high uncertainty of georeactor antineutrino spectrum. Such a nonstationary regime generates yet another, quite deep and nontrivial problem, i.e., the so-called problem of nonequilibrium neutrino spectra. Rejection of this problem can lead to serious errors in fitting of the experimental neutrino spectra. Below we consider this in more detail.

***On reasons and the degree of nonequilibrium of antineutrino spectra in KamLAND experiments.*** To describe the nuclear fuel antineutrino radiation the nuclide equilibrium concentrations of fission-product mixture and, accordingly, equilibrium antineutrino spectra obtained for hypothetical infinite irradiation time, which provides an secular equilibrium of all without exception fission products, are traditionally used as zeroth-order approximation.

On the other hand, it is obvious that the equilibrium approximation is not true for the non-stationary nuclear georeactor (Figure 14). Therefore, there is a question, how the strategy of calculation of effective neutrino spectra changes in this case and, in particular, how the resulting neutrino spectrum shape changes due to taking into account nonequilibrium neutrino spectra instead of equilibrium neutrino spectra, which are used for stationary nuclear reactors.

As is known, the direct summation method of $\beta$, $\nu$-spectra of individual nuclides operating [42, 43], of which the fission-product mixture consists at the specific modes of fuel irradiation in a nuclear reactor, and proper total effective $\beta$, $\nu$-spectrum of the nuclear system $k$

$$\rho^k(E) = \sum_j \lambda_j N_j^k \rho_j(E), \tag{23}$$

are used as the calculation algorithm when passing from the $\beta$-spectrum to the antineutrino spectrum. Here $\lambda_j$ is the decay probability of $j$th nuclide; $N_j$ is the number of nuclei of $j$th nuclide in the system $k$; $\rho_j(E)$ is the total $\beta,\nu$-spectrum of $j$th nuclide normalized to the nuclear decay:

$$\rho_j(E) = \frac{K_j}{\sum_i v_{i,j}} \sum_i v_{i,j} \rho_{i,j}(E). \tag{24}$$

where $\lambda_j N_j$ is the activity of $j$th nuclide depending on irradiation mode (fuel initial composition, neutron flux density, fuel burnup and other parameters influencing on accumulation of each $j$th nuclide); $K_j$ is the branching factor for $\beta-$ decay channel, i.e., the number of $\beta$–particles per decay; $\rho_{i,j}(E)$ is the partial $\beta$-transition spectrum of $j$th nuclide; $v_{i,j}$ is the $\beta$-transition intensity of $j$th nuclide.

A priori knowledge (based on calculation or experimental estimation) of the initial concentration $N_j(t)$ of $j$th fissionable actinoid and the one-group integral neutron flux density $\Phi$ makes it possible to determine the accumulation of one or another $j$th fission product in the reactor core by solving the following system of kinetic equations[*]

---

[*] Note that the index $i$ changes in the range $1 \leq I \leq p$, and according to the known Russian catalogue of radioactive nuclides total $\beta$, $\nu$–spectra [42] the index $j$ changes in the range $1 \leq j$ 1028, i.e., the talk turns to necessity to solve the enormous system of enchained differential equations. The method of solution of the system (25) based on the derivation of recurrence relations for $N_j(t)$ is in detail described in [53] and realized as the AFPA program package *(Accumulation of Fission Products and Actinides)* in terms of FORTRAN–IV.

$$\frac{dN_j}{dt} = -\lambda_j N_j - <\sigma_c>_j \Phi N_j + \sum_{i=1}^{p} <\gamma>_{ij} <\sigma_f>_i \Phi N_i +$$

$$+ \sum_{m=1}^{j-1} \lambda_{mj} N_m + \sum_{m=1}^{j-1} <\sigma_c>_{mj} \Phi N_m , \qquad (25)$$

which describes time change of the $j$th nuclide concentration $N_i(t)$ in fission-product mixture at the initial condition $N_j(0) = N_{0j}$ at the time $t$ linked with the activity of the $j$th nuclide in the following way $Q_j(t) = \lambda_j \cdot N_j(t)$; index "$m$" relates to precursor nucleus, $m < j$; $<\gamma>_{ij}$ is the independent yield of the $j$th nuclide due to the fuel $i$th component fission averaged over the effective neutron spectrum; $\lambda_{mj}$ is the decay probability of $m$th nuclide into the $j$th nuclide due to $\beta^-$, $\beta^+$–decay, electron capture, isomeric transition etc.; $<\sigma_f>_i$ is the one-group fission cross-section for the $i$th fissionable actinoid; $<\sigma_c>_i$ is the one-group $(n, \gamma)$, $(n, 2n)$ reaction cross-section for $i$th nuclide; $<\sigma_c>_{mj}$ is the $(n, \gamma)$, $(n, 2n)$ reaction cross-section for $m$th nucleus with $j$th nucleus formation.

Finding of time dependence of nuclide concentration of fission-product mixture (see (25)) is a sufficiently labor-consuming problem, whose solution depends on the specific conditions of fuel irradiation, i.e., the time dependence of neutron flux density, neutron flux spectral composition and also fuel initial composition [42, 43]. Therefore the exact solution of the kinetic system of equations (25) becomes practically inaccessible in the study of fission-product build-up in nonstationary nuclear reactors with nonconstant or in general unknown parameters. All above-said applies in full measure to the nonstationary nuclear georeactor, the thermal power evolution of which is shown in Figure 14. Moreover, in this case a situation is aggravated because, as was noted above, the $^{239}$Pu fission cross-section (the main active component of georeactor nuclear fuel) is strongly nonlinear magnitude, which grows under the power law in the temperature range 3000-5000 K (Figure 3), which is typical for the near-surface layers of the Earth solid core.

How much is better, when the reactor is stationary. In this case, the left-hand sides of the system of the equations (25) can be set equal to zero, and the system oneself is transformed into the system of linear algebraic equalizations, whose solution (the so-called equilibrium nuclide concentrations of fission product mixture) does not depend on initial conditions and irradiation time. The obtained equilibrium nuclide concentrations of fission product mixture make it possible (according to Eq.(23) to determine equilibrium partial and total neutrino spectra, which are usually used for description of effective neutrino spectra of stationary neutrino sources and, in particular, stationary nuclear reactors.

Finally, returning to the analysis of concrete KamLAND neutrino spectra, it is necessary to state that within the framework of georeactor hypothesis an integral fraction of reactor geoneutrino is sufficiently great and makes up almost the half of integral fraction of antineutrinos produced by all Japanese reactors in the KamLAND-experiment (see Figures 4 and 13). This means that the non-equilibrium property inherent to the reactor geoneutrino spectrum is not only delegated to the KamLAND experimental neutrino spectrum to a considerable extent, but plays a dominant role in natural distortion of this spectrum with respect to the KamLAND theoretical neutrino spectrum, which is based on the ideology of equilibrium neutrino spectra.

The question arises, to which degree this non-equilibrium influences on the effective neutrino spectrum shape in the general case and, for example, in KamLAND-experiments. As is shown in numerous test experiments related to the nuclear fuel irradiation under unstable conditions [42, 53], the non-equilibrium effect manifests itself as the observed distortion of some pieces on the nonequilibrium effective spectrum (with respect to an analogical equilibrium neutrino spectrum), whose location in spectrum energy coordinates is completely predetermined by the time dependence of neutron flux density and neutron flux spectral composition and fuel initial composition [42, 53].

On the other hand, the analysis of experimental KamLAND-data obtained over the years of 2002-2004, 2002-2007 and 2002-2009 shows that in all considered cases the number of recorded events in the fifth ($E_5$=2.8 MeV) and ninth ($E_9$=4.5 MeV) energy bins of experimental neutrino KamLAND-spectra considerably differs from similar data obtained by fitting or, in other words, theoretical equilibrium neutrino KamLAND-spectra (see Figure 1 in [4], Figure 1 in [5] and also Figures 4 and 13). At the same time, the number of recorded events in the fifth bin always is substantially less than in the fifth bin of theoretical neutrino KamLAND-spectrum, whereas reversed situation is observed in the ninth bin. This is the so-called problem of 5 and 9 energy bins of neutrino KamLAND-spectra, which, in our opinion, is caused not only by detection statistics, but is, for the most part, the manifestation of substantial non-equilibrium of neutrino spectrums. According to [42,43], the power of non-equilibrium effect, i.e., the difference between calculated equilibrium neutrino spectra and corresponding experimental non-equilibrium neutrino spectra, can attain 10-15 % and more.

At last, it is necessary to remind that in this paper all theoretical neutrino spectra (Figures 2, 4, 9 and 13) are built using the ideology of equilibrium spectra. Within the framework of georeactor hypothesis such an approach is reasonable, because the possible high degree of non-equilibrium of experimental neutrino spectra, which manifests itself, for example, as so-called problem of the 5 and 9 energy bins of neutrino KamLAND-spectra, is effectively compensated by even if high, but reasoned degree of uncertainty of theoretical neutrino spectra.

***Geological (magnetic) time-scale and time evolution of the nuclear georeactor heat power.*** Within the framework of alternative hypothesis we have obtained the time evolution of average georeactor heat power over the period of 2002-2009 (Figure 14), which shows that the average georeactor heat power $W_t$ sharply falls from 50 TW till 5-7 TW in KamLAND-experiments over the periods of 2002-2004, 2005-2007 and 2008-2009. Here a natural question arises: "What does such a dynamics reflect or what physical mechanism causes such a dynamics?". In other words, is it the display of a certain unknown physics or, vice versa, trivial consequence of the "…happily guessed rules of calculations not reflecting the veritable nature of things" [54]. Below we try to give the simple and physically clear interpretation of possible fundamental mechanism and its influence on the time dynamics of the nuclear georeactor heat power.

It is known that in spite of a long history the nature of the energy source maintaining a convection in the liquid core of the Earth or, more exactly, the mechanism of the magneto-hydrodynamic dynamo (MHD) generating the magnetic field of the Earth still has no clear and unambiguous physical interpretation [26, 55]. The problem is aggravated because of the fact that none of candidates for an energy source of the Earth magnetic-field [55] (secular cooling due to the heat transfer from the core to the mantle, internal heating by radiogenic isotopes, e.g., $^{40}K$, latent heat due to the inner core solidification, compositional buoyancy due to the ejection of light element at the inner core surface) can not in principle explain one of the most remarkable phenomena in solar-terrestrial physics, which consists in strong (negative) correlation between the temporal variations of magnetic flux in the tachocline zone (the bottom of the Sun convective zone) [56,57] and the Earth magnetic field (*Y*-component)[*] [58] (Figure 16).

At the same time, supposing that the transversal (radial) surface area of tachocline zone, through which a magnetic flux passes, is constant in the first approximation, we can consider that magnetic flux variations describe also the temporal variations of magnetic field in the solar tachocline zone. Thus, Figure 16 demonstrates simultaneously the mirror correlation of the temporal variations of magnetic field of the solar tachocline zone and the Earth magnetic field (*Y*-component).

---

[*] Note that the strong (negative) correlation between the temporal variations of magnetic flux in the tachocline zone and the Earth magnetic field (*Y*-component) will be observed (Figure 16) only for experimental data obtained at that observatories where the temporal variations of declination ($\delta D/\delta t$) or the closely associated east component ($\delta Y/\delta t$) are directly proportional to the westward drift of magnetic features [59]. This condition is very important for understanding of physical nature of indicated above correlation, so far as it is known that just motions of the top layers of the Earth's core are responsible for most magnetic variations and, in particular, for the westward drift of magnetic features seen on the Earth's surface on the decade time scale. Europe and Australia are geographical places, where this condition is fulfilled (see Figure 2 in [59]).

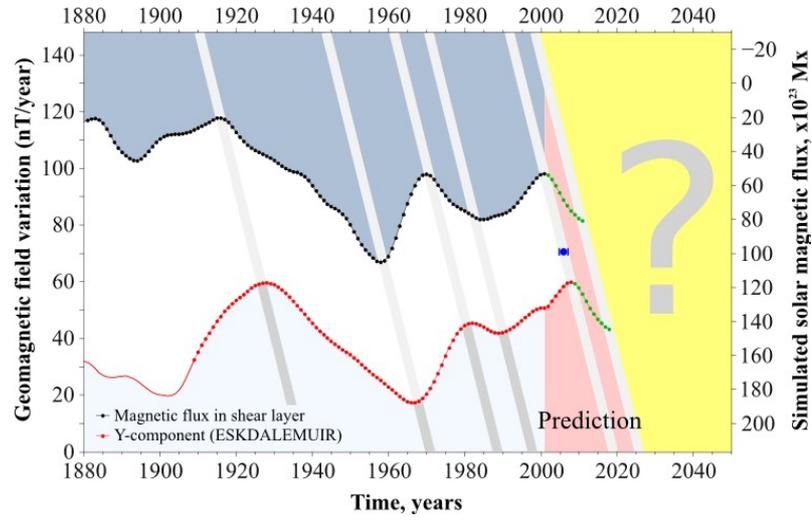

**Figure 16.** Time evolution of the variations of magnetic flux in the bottom (tachocline zone) of the solar convective zone ((black dotted line, see Fig.17), variations of the nuclear georeactor thermal power (blue point with bars) and geomagnetic field secular variations (*Y*-component, nT/year, red dotted line) [58] and prediction (green dotted line) [58]. All curves are smoothed by sliding intervals in 5 and 11 years. The pink area is the prediction region.

To obtain such a obvious correlation we used the moving-average process. In particular, to average the sequence $\{x_n\}$ and obtain the averaged sequence $\left[\langle x \rangle_n^N\right]$ we used the following expression

$$\langle x \rangle_n^N = \frac{1}{N}\left(\sum_{i=1}^{(N-1)/2}(x_{n-i}+x_{n+i})+x_n\right), \quad where \quad N=2k-1\geq 3 \ . \tag{26}$$

where *k* is the positive integer.

The smoothed curve of variations of magnetic flux of the solar tachocline zone, which is shown in Fig, 17c (black dotted line), demonstrates the result of such an averaging of the initial curve (red fill area in Figure 17c) by the two sliding intervals in $N_1=5$ and $N_1=11$ years. Physical sense of such a double averaging consists in the "soft" removing of influence of the 11-year solar period and obtaining of the so-called amplitude-modulated representation of magnetic flux variations of the solar tachocline zone.

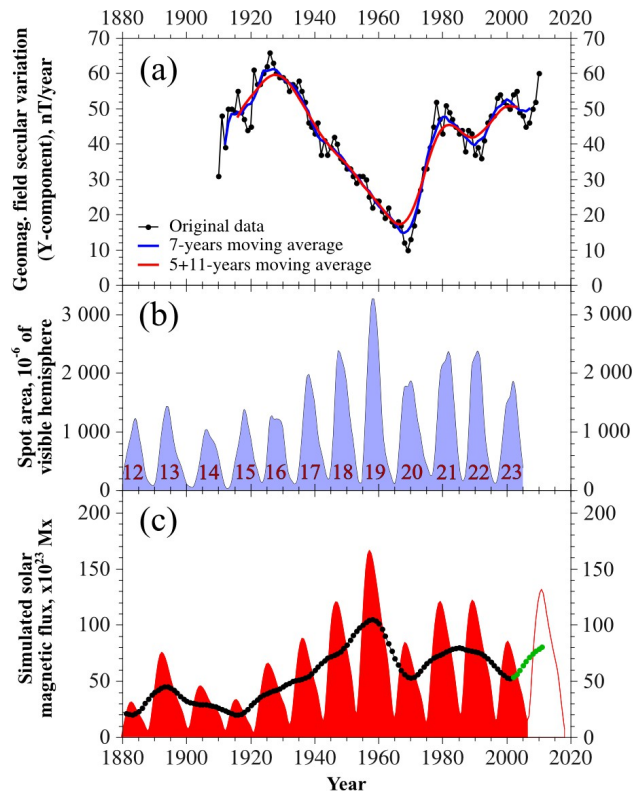

**Figure 17.** Time evolution of (a) geomagnetic secular variations (Y-component, nT/year) [58], (b) observed sunspot area for cycles 12-23 [57], (c) intergral from $0°-45°$ latitude of simulate-ion toroidal magnetic flux in bottom shear layer (red shadow zone) for cycles 12–23 [57], plus forecast for cycle 24 [57]. Black curve are smoothed by sliding intervals in 5 and 11 years. The green dotted line is the prediction region.

From Fig. 16 follows that the degree of averaging effect on time sample of the variations of solar magnetic field and the Earth magnetic field (Y-component) sharply differs in the degree of their smoothing. In our opinion, this is caused by the considerable delay (10-12 years) of the variations of terrestrial magnetic field (Y-component) with respect to the solar magnetic field variations, during of which intermediate deep terrestrial processes controlled by the solar power pacemaker not only activate and maintain the Earth magnetic field, but appreciably smooth some fine "details" of physical influence of solar power pacemaker (Fig. 17). Below we consider the physical mechanism of the one possible intermediate deep terrestrial process by virtue of which a future candidate for an energy source of the Earth magnetic field must play not only the role of a natural trigger of solar-terrestrial connection, but also directly generate the solar-terrestrial magnetic correlation by its own participation.

We assume that the temporal variation of soliton-like nuclear georeactor heat power can be a global energy cause of such a fundamental geophysical phenomenon as the variations of terrestrial magnetic field (Y-component). If this is truth, it is obvious that exactly the variations of solar magnetic field must "govern" or, in other words, to be reason of the temporal variations of nuclear georeactor thermal power, which, according to our hypothesis, is the energy source of the Earth magnetic field and its variations, respectively. One of possible mechanisms generating strong (negative) correlation between the terrestrial magnetic field and solar magnetic field (see Fig. 16) can be so-called axion mechanism of solar dynamo-geodynamo connection [60].

The essence of this mechanism is that the total energy of $^{57}$Fe-axions produced mainly in the Sun core is modulated at first by the magnetic field of the solar convective zone (due to the inverse coherent Primakoff effect [60]) and after that is resonance absorbed by $^{57}$Fe in the Earth core. In other words, the higher the solar magnetic field, the greater the number of axoins is converted (by the inverse Primakoff effect) into $\gamma$-quanta, the smaller the number of axoins reach the Earth and are absorbed in the Earth core, and vice versa.

It results in the fact that the variations of $^{57}$Fe-axion intensity play the role not only of heat source, which changes the temperature of the Earth core, but also the modulator of nuclear georeactor thermal power, because the medium temperature in the range 2500-6000 K modulates the value of the $^{239}$Pu fission cross-section (Fig. 3). In other words, the solar axion's mechanism not only explains the nature of heat source in Earth liquid core, which plays the role of the modulator of nuclear georeactor thermal power, but in a natural way explains the cause of experimentally observed strong negative correlation (Fig. 16) between the magnetic field of the solar convective zone and the Earth magnetic field (Y-component).

From Figure 16 it follows that the variation of the Earth magnetic field (Y-component) lags behind the variation of the solar magnetic field about 10-12 years. On the other hand, it is known that a magnetic signal predetermined by an extremum of drift velocity of eccentric dipole of the Earth core lags and therefore is detected on the Earth surface 5-7 years late [61, 62].

Within the framework of georeactor hypothesis this means that the temporal variation of magnetic field energy $W_{SE}$ on the Earth surface also has the delay of 5-7 years with respect to the temporal variation of magnetic field energy $W_{core}$ of the Earth liquid core which, in its turn, forms with the delay with respect to the temporal variation of the nuclear georeactor thermal power $W$. At the same time, because $W_{core} \sim B^2$, where $\vec{B}$ is the magnetic inductance vector, we can obtain from the obvious relation $W_{core} \sim W$ that

$$B \sim \sqrt{W} \ . \tag{27}$$

On the other hand, we have found the sampling of values for the nuclear georeactor thermal power (Table 1), which is obtained by the experimental KamLAND data handling over the years of 2002-2009. This sampling contains the three averaged values: 50.2 TW over the years of 2002-2004, 21.1 TW over the years of 2005-2007 and 7.3 TW over the years of 2008-2009. Due to delay of temporal variation of the nuclear georeactor thermal power with respect to the variation of the Earth magnetic field (Y-component) it is obvious that to smooth over the influence of background processes (the variations of the Earth liquid core temperature, nuclear fuel composition etc.) accompanying geodynamo operation, it is necessary to average the sampling of values of the nuclear georeactor thermal power by a sliding time interval, whose length is of order delay time, i.e., N=5-7 years. It is clear that, if to use a maximum possible sliding interval with $N=7$, this sampling composed of the 7 virtual annual values over the years of 2002-2009 will be transformed into the sampling which contains only one term characterizing the nuclear georeactor average thermal power

$$\langle W \rangle_4^7 \sim 30 \quad TW \ . \tag{28}$$

It is obvious that this single term of new sampling corresponds to the year 2006. If according to the solar axion mechanism to assume the existence of strong (negative) correlation between the variation of the Earth magnetic-field and the value $\sqrt{W}$ (see Eq. (27) and also to take into account the delay time (5-7 years) of variation of $\sqrt{W}$ with respect to the variation of the solar magnetic-field, it is easy to find the coordinates of nuclear georeactor in Fig. 16. The value of $\sqrt{W}$ is at the intersection of vertical line $t$=2006 and the slanting grey line passing through the extreme point ($t$=2001) on the curve of variation of the solar magnetic field. Recall that the slope of grey straight line in Fig. 16 is the effect of delay under the conditions of strong (negative) correlation of the solar magnetic field and the Earth magnetic field (Y-component).

Note that all the future measurements of annual variations of neutrino flux in the KamLAND and Borexino experiments will generate new theoretical data describing the variations of nuclear georeactor thermal power. If these variations smoothed by moving-average process will correlate with the variations of the solar magnetic field and variations of the Earth magnetic field (Y-component), the georeactor hypothesis will obtain weighty indirect confirmation.

Finally, we would like to remind that within the framework of georeactor hypothesis a forecast of behavior of considered above fundamental geophysical processes, which have the common nature (the temporal variation of magnetic field of the solar tachocline zone), is possible only up to corresponding event horizon predetermined by delay time of variation of the nuclear georeactor thermal power (5-7 years) or magnetic Y-field of the Earth (10-12 years) with respect the magnetic field of the solar tachocline zone. It is obvious that such a delay effect makes it possible reliably to predict the behavior of the Earth magnetic field (Y-component) by experimental observation of georeactor antineutrino, whose variations characterize the variations of nuclear georeactor thermal power.

## 7. Conclusion

We should note that although the nuclear georeactor hypothesis which we used for the interpretation of KamLAND-experiment seems to be very effective, it can be considered only as a possible alternative variant for describing the KamLAND experimental data. Only direct measurements of the geoantineutrino spectrum in the energy range >3.4 MeV in the future underground or submarine experiments will finally settle the problem of the existence of a natural georeactor and will make it possible to determine the "true" values of the reactor antineutrino oscillation parameters. At the same time, just solution of the direct and the inverse problems of the remote neutrino-diagnostics for the intra-terrestrial processes which is essential to obtain the pure geoantineutrino spectrum and to determine correctly the radial profile of the $\beta$-sources in the Earth's interior [43, 63] will undoubtedly help to settle the problem of the existence of a natural nuclear reactor on the boundary of the liquid and solid phases of the Earth's core.

In the second part of this paper [46] we will consider some properties of those fundamental geophysical phenomena, which must be observed directly under terrestrial conditions, if a georeactor hypothesis is true and the nuclear georeactor exists.